\documentclass[aps,prc,amsfonts,preprintnumbers,superscriptaddress,showpacs,nofootinbib]{revtex4}
\pdfoutput=1
\usepackage{graphics}
\usepackage{amssymb}
\usepackage{psfrag}
\usepackage{epsfig}
\usepackage{epsf}
\usepackage{float}
\usepackage{slashed}
\usepackage{amsmath,amsthm}
\allowdisplaybreaks

\newcommand{\fet}[1]{\mbox{\boldmath $#1$}}

\newcommand{\beq}{\begin{equation}}
\newcommand{\eeq}{\end{equation}}
\newcommand{\beqa}{\begin{eqnarray}}
\newcommand{\eeqa}{\end{eqnarray}}

\begin{document}

\title{Low-energy theorems for nucleon-nucleon scattering at unphysical pion masses}

\author{V. Baru}
\affiliation{Institut f\"ur Theoretische Physik II, Ruhr-Universit\"at Bochum,
  D-44780 Bochum, Germany}
\affiliation{Institute for Theoretical and Experimental Physics, B. Cheremushkinskaya 25, 117218 Moscow, Russia}

\author{E.~Epelbaum}
\affiliation{Institut f\"ur Theoretische Physik II, Ruhr-Universit\"at Bochum,
  D-44780 Bochum, Germany}

\author{A. A. Filin}
\affiliation{Institut f\"ur Theoretische Physik II, Ruhr-Universit\"at Bochum,
  D-44780 Bochum, Germany}

\author{J.~Gegelia}
\affiliation{Institut f\"ur Theoretische Physik II, Ruhr-Universit\"at Bochum,
  D-44780 Bochum, Germany}
\affiliation{Tbilisi State University, 0186 Tbilisi, Georgia}

\date{\today}

\begin{abstract}
The longest-range part of the nuclear force due to the one-pion exchange governs the energy dependence
of the scattering amplitude in the near-threshold region and imposes correlations
between the coefficients in the effective range expansion. These correlations  may be regarded as
low-energy theorems and are known to hold to a high accuracy in the neutron-proton
$^3$S$_1$ partial wave. We generalize the low-energy theorems to the case of unphysical pion masses and
provide results for the correlations between the coefficients in the effective range expansion
in this partial wave  for pion masses up to $M_\pi \sim 400\;\mbox{MeV}$. We discuss the implications
of our findings for the available and upcoming lattice-QCD simulations of two-nucleon observables.
\end{abstract}


\maketitle

\vspace{-0.2cm}

%%%%%%%%%%%%%%%%%%%%%%%%%%%%%%%%%%%%%%%%%%%%%%%%%%%%%%%%%%%%%%%%%%%%%%%%%%%%%%%%%
\section{Introduction}
\def\theequation{\arabic{section}.\arabic{equation}}
\label{sec:intro}

The dependence of the nuclear forces and, more generally, nuclear observables upon the fundamental parameters
of the Standard Model such as the light quark masses is one of the open problems in contemporary theoretical physics.
Being a fascinating question on its own, it has attracted considerable attention on the theory side
and has been shown to play important
role for various phenomena in nuclear,  astro- and particle physics. In particular, Bulgac et al.~have
speculated about an intriguing possibility of the appearance of a bound state in the $^3$P$_0$ neutron-proton partial wave
in the so-called chiral limit of Quantum Chromodynamics (QCD) corresponding to massless up and down quarks \cite{Bulgac:1997ji}.
Such a scenario would clearly have far reaching consequences for nuclear physics.
Another interesting observation was made in Ref.~\cite{Braaten:2003eu}, where the proximity
of the physical quark masses to the critical trajectory for   an infrared renormalization group limit cycle
of QCD was pointed out and conjectured to be responsible for the success of Efimov's program for describing
the three-nucleon problem, see also Ref.~\cite{Epelbaum:2006jc} for a follow-up study along this line.

The quark mass dependence of the nuclear force plays an important role
for constraining certain types of extensions of the Standard Model. In particular, many such theories
allow for the parameters of the Standard Model such as the light quark masses to vary over time.
The knowledge of the quark mass dependence of the nuclear force combined with the theory of Big Bang
nucleosynthesis (BBN) and the observed nuclear abundances allow one to constrain a possible
variation of quark masses at the time of BBN \cite{Bedaque:2010hr,Berengut:2013nh}.

Recently, this topic has experienced renewed interest
in connection
with the anthropic considerations in nuclear physics  \cite{Meissner:2014pma}. In particular, the second $0^+$ state of $^{12}$C,
the so-called Hoyle state, is well-known to play a crucial role for a formation of life-essential elements such
as  $^{12}$C and $^{16}$O in hot, old stars due to its closeness to the $^4$He--$^8$Be threshold \cite{Hoyle:1954zz}.
Changing its excitation energy of $\epsilon = 379.47 (18)$ keV, measured with respect to the triple-$\alpha$ threshold,
by more than $\sim 25\%$ was shown to strongly decrease the
production rate of either $^{12}$C or $^{16}$O
\cite{Oberhummer:2000mn,Schlattl:2003dy}. The Hoyle state  is,
therefore,
viewed as a promising
candidate to address the fine-tuning problem of the fundamental constants of Nature in
connection with the anthropic principle. The sensitivity of the energy difference $\epsilon$ to a variation of the light
quark masses was recently analyzed within an \emph{ab initio} framework of nuclear lattice simulations
\cite{Epelbaum:2012iu,Epelbaum:2013wla}. This theoretical approach
makes use of a discretized formulation of chiral effective field theory (EFT) combined with the Monte Carlo method to
perform Euclidean time evolution of an $A$-nucleon state, see Ref.~\cite{Lee:2008fa} for a review article.
Presently, by far the dominant
source of the theoretical uncertainty in
this calculation is related to the lack of knowledge of the quark mass dependence of the nuclear force
or, more precisely, of the nucleon-nucleon  (NN) S-wave scattering lengths.

While the ultimate answer to the question of quark mass dependence of hadronic observables is eventually to be provided
by lattice QCD, one can gain useful insights into this topic within the framework of chiral EFT.
During the past two decades, this theoretical
approach was developed into a powerful tool to derive nuclear two- and many-body forces and the corresponding
current operators in a systematically improvable  and model-independent way, see
Refs.~\cite{Epelbaum:2008ga,Machleidt:2011zz}
for recent review articles. It exploits
the symmetry and symmetry-breaking pattern of QCD to formulate an appropriate EFT in terms of pions
and nucleons (and possibly of the $\Delta$(1232) isobar) which are the relevant degrees of freedom for
low-energy nuclear physics in the non-strange sector. The resulting effective Lagrangian has been used to
analyze hadronic observables in the Goldstone-boson and single-nucleon sectors as well as nuclear forces
and the corresponding current operators by means of the chiral expansion, i.e.~a perturbative expansion
in powers of the pion mass $M_\pi$ and three-momenta of external particles. In particular, for the two-nucleon
force, this expansion has been pushed to next-to-next-to-next-to-next-to-leading order \cite{Epelbaum:2014sza,Entem:2014msa} and
demonstrated to provide very accurate description of low-energy NN scattering observables
and the deuteron properties \cite{Epelbaum:2014sza,Epelbaum:2014efa}.
Notice that while the number
of the relevant low-energy constants (LECs) in the effective Lagrangian
grows
with an increasing order of the calculation,
the large amount of available experimental data on
low-energy proton-proton and neutron-proton scattering observables
make their determination unproblematic.

Given its reliance on the spontaneously broken approximate chiral symmetry of QCD, chiral EFT
provides, at least in principle, a suitable theoretical framework to address quark mass dependence of
the nuclear forces in a systematic way, see
Refs.~\cite{Beane:2001bc,Beane:2002xf,Epelbaum:2002gb,Beane:2002vs,Epelbaum:2002gk,Braaten:2003eu,Epelbaum:2006jc,Soto:2011tb,Bedaque:2010hr,Berengut:2013nh,Epelbaum:2012iu,Epelbaum:2013wla,Barnea:2013uqa}
for recent studies along this line and
Refs.~\cite{Bulgac:1997ji,Flambaum:2007mj} for related calculations at the more phenomenological level.
Clearly, the major complication  which prevents one from being able to perform accurate chiral EFT calculations
of few- and many-nucleon systems at unphysical values of the quark masses is the lack of knowledge of the
corresponding LECs.
Quark mass dependence of the short-range interactions induced  by integrating out the momentum scale
associated with real pion production has been discussed in Ref.~\cite{Mondejar:2006yu}.
Further complications emerge due to the appearance of a finite cutoff in the calculations.
This feature is unavoidable if the one-pion
exchange potential is iterated in the non-relativistic
Lippmann-Schwinger equation to all orders.
Such an approach relies on implicit renormalization which makes it difficult in practice to control the quark mass dependence of  short-range operators in a
systematic way.
Notice that the above-mentioned complications can be avoided by treating the exchange of pions in  perturbation
theory using e.g.~the approach proposed by Kaplan, Savage and Wise (KSW) \cite{Kaplan:1998we},  see Ref.~\cite{Beane:2002xf} for studies along this
line and Ref.~\cite{Soto:2011tb} for calculations using a closely related approach with dibaryon fields.
The perturbative treatment of the one-pion exchange within this framework was, however, shown to be
inadequate in spin-triplet channels of NN scattering \cite{Cohen:1998jr,Fleming:1999ee}.

Recently, an attempt was made to overcome the above mentioned difficulties associated with the unknown quark mass dependence
of the short-range part of the nuclear force by employing a resonance saturation hypothesis for contact NN operators
\cite{Berengut:2013nh}. It is well known that the LECs accompanying the NN contact interactions can be
understood at a semi-quantitative level  in terms of the exchange of heavy mesons in the sense of  resonance saturation
\cite{ressat}.
Using unitarized chiral perturbation theory and the information provided by lattice-QCD simulations
 to determine
the pole positions of the relevant resonances as functions of the quark masses and assuming the
validity of the resonance saturation picture at unphysical pion masses allows one to predict the quark
mass dependence of the contact operators and to carry out chiral extrapolations in the NN sector.
This strategy was followed in Ref.~\cite{Berengut:2013nh} up to  next-to-next-to-leading order
in chiral EFT within the Weinberg power counting scheme. It was, in particular, found that the deuteron is
likely to become less bound at larger values of the light quark mass.  These findings are in  good agreement with the
ones achieved in Ref.~\cite{Flambaum:2007mj} using a more phenomenological approach by considering
a set of Argonne models for NN interaction. A detailed comparison of these results with the
available earlier chiral EFT calculations both within the Weinberg and the KSW power counting
schemes can be found in Ref.~\cite{Berengut:2013nh}.
We emphasize, however, that  modelling  the $M_{\pi}$ dependence of the contact interaction,  as done in Ref.~\cite{Berengut:2013nh},
provides a
 substantial source of uncertainty   which  was not estimated in that work.

Finally, chiral extrapolations of the NN S-wave scattering lengths and the deuteron binding energy were considered
in Ref.~\cite{Epelbaum:2013ij} using the chiral EFT formulation of Ref.~\cite{Epelbaum:2012ua}.
In this approach, the leading-order (LO) NN scattering amplitude is obtained by solving the
three-dimensional integral equation introduced originally by Kadyshevsky \cite{Kadyshevsky:1967rs}. This equation
is an example of three-dimensional integral equations which satisfy relativistic elastic unitarity.
Clearly, the Kadyshevsky equation turns into the usual Lippmann-Schwinger equation upon taking the non-relativistic limit of
the two-nucleon propagator. An important feature of the Kadyshevsky equation for the LO
NN scattering amplitude is its renormalizability. In particular, \emph{all} ultraviolet divergences
generated by iterations can be explicitly absorbed into redefinition of the
NN derivative-less contact interaction. This feature allows one to remove the ultraviolet cutoff
when calculating the amplitude and to avoid the above mentioned complications emerging in the non-relativistic framework with a finite cutoff.
Assuming that the contributions of
the $M_\pi^2$ dependent short-range operators are suppressed relative to the
derivative-less, $M_\pi^2$ independent contact terms, the quark
mass dependence of the
S-wave phase shifts at LO emerges entirely from the propagator in the one-pion exchange potential.
The resulting parameter-free predictions in the spin-triplet neutron-proton channel
are in  very good agreement with the calculations reported in   Ref.~\cite{Berengut:2013nh}.
Also the $M_\pi$ dependence of the $^1$S$_0$ scattering length agrees qualitatively with the
one reported in that paper.

While all above-mentioned studies \cite{Flambaum:2007mj,Berengut:2013nh,Epelbaum:2013ij},
which are carried out using rather
different theoretical approaches,
predict less attraction in the deuteron channel for increasing values of
the light quark masses, the available lattice-QCD simulations seem to indicate  an opposite trend.
The first pioneering calculation of NN scattering in the framework of lattice QCD has been
carried out two decades ago by Fukujita et al.~\cite{Fukugita:1994ve} within the quenched approximation
and using pion masses of $M_\pi \simeq 550$ MeV and heavier. Ten years later, the first fully dynamical lattice
QCD calculation of the NN S-wave scattering length at pion masses of $M_\pi \simeq 350$ MeV, $490$ MeV and  $590$ MeV
were reported in Ref.~\cite{Beane:2006mx}. This study found smaller values
of the $^3$S$_1$ scattering length (corresponding to a stronger bound deuteron) but was inconclusive
of its sign. Notice further that similarly to the calculation
reported in Ref.~\cite{Fukugita:1994ve}, the authors of  Ref.~\cite{Beane:2006mx}
did not address  volume dependence. Recently, binding energies of light nuclei
have been calculated in 3-flavor QCD at  $M_\pi \simeq 390$ MeV  \cite{Beane:2011iw}
and $M_\pi \simeq 800$ MeV  \cite{Beane:2012vq}
as well as in  $2+1$-flavor QCD at $M_\pi \simeq 510$ MeV \cite{Yamazaki:2012hi}, see also Ref.~\cite{Yamazaki:2009ua} for  a
quenched QCD calculation of the same observables. These fully dynamical lattice QCD calculations
found a considerably stronger bound deuteron at these large pion masses, see section \ref{sec:LatticeQCD} for more details.
Notice further that the NPLQCD Collaboration also extracted the values of the
effective range and the first shape coefficient at the
pion mass of $M_\pi \simeq 800$ MeV \cite{Beane:2013br}.
On the other hand, the HAL QCD Collaboration found no bound state in the NN $^3$S$_1$-$^3$D$_1$
channel by carrying out 3-flavor QCD simulations at pseudoscalar meson masses of $469 \ldots 1171$ MeV
and employing a different approach by making use of the two-nucleon
potential at the intermediate step of the calculation \cite{Inoue:2011ai}. Very recently,
Yamazaki et al.~have performed $2+1$-flavor QCD calculations of light nuclei at pion mass
of $M_\pi \simeq 300$ MeV  \cite{Yamazaki:2015asa}. They found the deuteron to be even stronger bound at $M_\pi = 300$ MeV
compared to their earlier study \cite{Yamazaki:2009ua}   at $M_\pi \simeq 510$ MeV. We further emphasize
that the preliminary results of the NPLQCD Collaboration at $M_\pi \simeq 430$ MeV  \cite{Beane:Prelim} also suggest
the deuteron to be stronger bound as compared to the physical case.
The current situation with the lattice-QCD results obtained by different
groups is thus not completely clear although there seems to be a common trend towards
stronger bound deuteron and other light nuclei at values of the light quark masses larger than the physical
ones.

In this paper we address the quark mass dependence of low-energy NN scattering
observables from the point of view of low-energy theorems (LETs). These theorems establish
model-independent relations
between the coefficients in the effective range expansion (ERE) which are determined by the
long-range part of the nuclear force governed by the pion exchange. We  generalize the LETs
to unphysical pion masses and discuss their accuracy and the validity range with
respect to the variation in $M_\pi$.  We apply the LETs to provide relations between the coefficients in
the ERE in the $^3$S$_1$ partial wave for pion masses up to $M_\pi \simeq 400$ MeV.
Our results open the way for nontrivial consistency checks of the existing and upcoming
lattice-QCD calculations (see also Ref.~\cite{Barnea:2013uqa} for a related discussion).
As an example of a possible application, we employ
a linear with respect to $M_\pi$ extrapolation of the $^3$S$_1$ effective range suggested
in Ref.~\cite{Beane:2013br} to predict the
resulting $M_\pi$ dependence of the scattering length, shape parameters and the deuteron binding
energy. The obtained results agree well with the lattice-QCD calculations reported in
Refs.~\cite{Beane:2011iw,Yamazaki:2012hi,Beane:Prelim}. We also use the LETs to predict
the values of the
scattering length, effective range and the shape parameters using the deuteron binding
energies calculated at $M_\pi = 300$ MeV \cite{Yamazaki:2015asa}, $M_\pi = 390$ MeV \cite{Beane:2011iw} and
$M_\pi = 430$ MeV \cite{Beane:Prelim} which can be tested on the lattice by calculating
the phase shifts.

Our paper is organized as follows. In section \ref{sec:LETphys} we discuss in detail the
meaning of the LETs and provide results at the physical value of the pion mass in the
$^1$S$_0$ and $^3$S$_1$ channels. A generalization of the LETs to unphysical values
of the pion mass in the $^3$S$_1$ partial wave is considered in section \ref{sec:LET}.
Implications of these findings for lattice-QCD results are addressed in section \ref{sec:LatticeQCD}.
The main results of our work are summarized in section \ref{sec:Summary}.

%%%%%%%%%%%%%%%%%%%%%%%%%%%%%%%%%%%%%%%%%%%%%%%%%%%%%%%%%%%%%%%%%%%%%%%%%%%%%%%%%
\section{Low-energy theorems for the physical pion mass}
\def\theequation{\arabic{section}.\arabic{equation}}
\label{sec:LETphys}

Long-range interactions are responsible for the near-threshold
left-hand singularities of the partial wave scattering amplitude  and control its
energy dependence
 \cite{Cohen:1998jr,Steele:1998zc}.
In particular,
they impose correlations between the coefficients
in the effective range expansion which can be regarded as low-energy
theorems. In the following, we discuss in some detail the meaning of
the LETs following the lines of
Refs.~\cite{Epelbaum:2009sd,Epelbaum:2009zz,Epelbaum:2010nr}
and using the framework
of non-relativistic quantum mechanics which is appropriate for
analyzing low-energy NN scattering. For a related discussion on
reconstructing the scattering amplitude in the physical region
based on the discontinuities across the left-hand cuts and employing the
unitarity constraints see Refs.~\cite{Gasparyan:2012km,Oller:2014uxa}
and references therein.

\begin{figure}[tb]
\vskip 1 true cm
\includegraphics[width=1.0\textwidth,keepaspectratio,angle=0,clip]{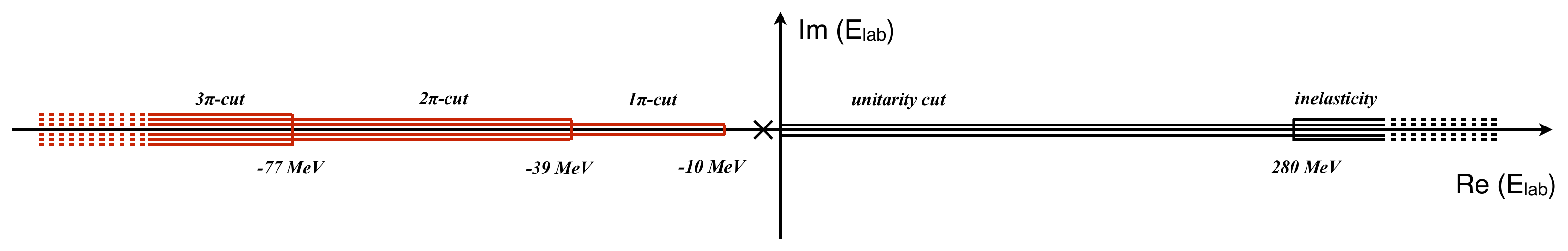}
    \caption{Cut structure of the partial-wave nucleon-nucleon scattering amplitude in
      the complex energy plane in the absence of electromagnetic
      interactions. The cross refers to a possible bound state.
\label{fig:cuts}
 }
\end{figure}

Consider two non-relativistic particles of mass $m$ interacting via some short-range potential
$V$. The corresponding $S$-matrix for an uncoupled channel with the orbital angular momentum $l$
is parametrized in terms of a single phase shift $\delta_l$ and can be written in terms of the $T$-matrix as
\beq
\label{Tnorm}
S_l = e^{2 i \delta_l (k) } = 1 - i \left( \frac{k m}{8 \pi^2} \right) T_l (k)\,,
\eeq
where $k$ denotes the scattering momentum in the center-of-mass system
(CMS). In the complex energy
plane, the partial-wave scattering amplitude and thus also the $T$-matrix
possess a so-called unitarity cut, a kinematic singularity due to two-body unitarity.
The unitarity cut starts from the branch point at the threshold ($E=0$) and goes to
positive infinity.  Furthermore, there are singularities associated with the interaction mechanism
and located at the negative real axis.
In particular, in the case of the Yukawa potential ($\sim \exp (-M r )/r$) corresponding to an exchange of
a meson of mass $M$, the amplitude has a left-hand cut starting at $k^2 = -M^2/4$.
Bound and virtual states reside as poles at the negative real axis ($k = i | k |$ and $k = -i | k |$ for
bound- and virtual-state poles, respectively) while resonances show up as poles at complex energies.
The cut structure of the nucleon-nucleon scattering amplitude in the
absence of long-range electromagnetic potentials is visualized in
Fig.~\ref{fig:cuts}.

It is useful to express the $T$-matrix
in terms of the so-called effective range function $F_l (k) \equiv k^{2l+1} {\rm cot} \delta_l (k)$ via
\beq
\label{tmat}
T_l (k) = -\frac{16 \pi^2}{m} \frac{k^{2l}}{F_l (k) - i k^{2l+1}} \,.
\eeq
Contrary to the scattering amplitude, the effective range function does not possess the kinematic
unitarity cut and can be shown to be a  real meromorphic (i.e. analytic except for poles) function of $k^2$ near the origin $k=0$ for non-singular
potentials of a finite range \cite{Blatt:49,Bethe:49}. It can, therefore, be
Taylor-expanded about the origin leading to the well-known effective
range expansion (ERE) which has the form
\begin{equation}
\label{ere}
 k^{2l+1} {\rm cot} \delta_l (k) = - \frac{1}{a} + \frac{1}{2} r k^2 +
 v_2 k^4 +  v_3 k^6 +  v_4 k^8 + \ldots\,,
\end{equation}
where $a$ and $r$ refer to the scattering length and the effective
range, respectively, while $v_i$ denote the so-called shape
parameters. The convergence radius of the ERE
is bounded from above by the lowest-lying left-hand singularity associated with the
potential. In particular, given that the longest-range part of the
strong nuclear force is due to the one-pion-exchange potential (OPEP), the
ERE for NN scattering is expected to
converge for energies up to $ | E_{\rm lab} | \sim M_\pi^2/(2 m_N)  =
10.5$ MeV. Notice that the actual convergence range of the ERE might be smaller
if the effective range function possesses poles corresponding to
zeros of the scattering amplitude  whose positions are determined by the strength
of the interaction. Such a situation emerges, for
example, when phase shifts change the sign. The problem of decreasing the
 applicability range of the ERE can be easily
avoided with if  the Taylor expansion is replaced by e.g.~Pade approximants, see
Ref.~\cite{Midya:2015eta} for a related discussion.
Furthermore, we
emphasize that a
generalization of the above considerations to coupled channels such as e.g. the
$^3$S$_1$--$^3$D$_1$ channel in neutron-proton scattering is straightforward and
amounts mainly to replacing the amplitude by a $2 \times 2$ matrix,
see Ref.~\cite{PavonValderrama:2005ku} for more details.

The framework of the ERE can be generalized to the case in which the potential is
given by a sum of  long-range ($r_L \sim
M_L^{-1}$) and short-range ($r_S \sim M_S^{-1} \ll M_L^{-1}$) potentials, $V_L$ and $V_S$,
respectively. Following van Haeringen and Kok \cite{vanHaeringen:1981pb}, one
can define the \emph{modified} effective range function $F_l^M$ via
\beq
\label{mere}
F_l^M (k^2) \equiv R_l^L (k) + \frac{k^{2l+1}}{|f_l^L (k)|} \cot [\delta_l
(k) - \delta_l^L (k)]\,.
\eeq
In this equation,
$f_l^L (k)$ denotes the
Jost function defined according to $f_l^L (k) \equiv f_l^L (k, r) \big|_{r = 0}$
with $f_l^L (k, r)$ being the Jost solution of the Schr\"odinger equation corresponding to the potential $V_L$,
i.e.~the particular solution that fulfils
\beq
\lim_{r \to \infty} e^{- i k r} f_l (k, \, r) = 1\,.
\eeq
Further, $\delta_l^L (k)$ denotes the phase shift associated with the potential $V_L$
and the quantity $R_l^L (k)$ can be computed from  $f_l^L (k, r)$ as follows:
\beq
R_l^L (k) = \left( - \frac{i k}{2} \right)^l \frac{1}{l!} \, \lim_{r \to 0} \left[\frac{d^{2l+1}}{dr^{2l+1}}
\, r^l\frac{f_l^L (k, \, r) }{f_l^L (k)} \right] \,.
\eeq
Here and in what follows, the superscript ``L'' indicates that the
quantity of interest can be calculated solely from the
long-range part of the potential $V_L$.
As proven in Ref.~\cite{vanHaeringen:1981pb}, the modified effective range function $F_l^M (k^2)$ does not contain the left-hand singularities
associated with the long-range potential and reduces, per construction, to the  ordinary effective range function
$F_l (k^2)$ in the case of $V_L=0$.
It is a real meromorphic function in a much larger region set by  $r_S^{-1}$ as  compared
to $F_l (k^2)$. In particular, for Yukawa-type potentials, the region
in which the modified effective range function is meromorphic is set
by $| k | < M_S/2$. Similarly to the ERE, one can Taylor expand the
function $F_l^M (k^2)$ near the origin via
\begin{equation}\label{MEREexp}
F_l^M (k^2) = - \frac{1}{a^M} + \frac{1}{2} r^M k^2 +
 v_2^M k^4 +  v_3^M k^6 +  v_4^M k^8 + \ldots\,.
\end{equation}
This expansion is referred to as the modified effective range
expansion (MERE).

The most frequently used application of the above framework concerns
proton-proton scattering. In that case, the long-range interaction  is
due to the Coulomb potential $V_L(r)  = \alpha/r$ with $\alpha$ being
the fine structure constant. The left-hand cut
starts directly at the threshold so that the ERE has zero radius of
convergence. Notice further that the  Jost solution and, consequently, the function $R_l^L (k)$
can be calculated analytically for the case of the Coulomb potential.
For example, for $l=0$ and the repulsive Coulomb potential,
the modified effective range function takes the following well-known form:
\beq
F_0^M (k^2) \equiv F_C  (k^2)  = C_0^2 (\eta ) \, k \, \cot[\delta
(k) - \delta^C (k)] + 2 k \, \eta  \, h (\eta  )\,,
\eeq
where the Coulomb phase shift is $\delta^C \equiv \arg \, \Gamma (1 + i \eta )$
and the quantity $\eta$ is given by
\beq
\eta = \frac{m}{2 k} \alpha  \,.
\eeq
Further,  the functions $C_0^2 (\eta )$ (the Sommerfeld factor) and $h (\eta )$ read
\beq
C_0^2 (\eta ) = \frac{2 \pi \eta }{e^{2 \pi \eta } - 1} \,,  \quad  \quad  \mbox{and} \quad  \quad
h (\eta ) = {\rm Re} \Big[ \Psi ( i \eta  ) \Big] - \ln (\eta  ) \,.
\eeq
Here, $\Psi (z) \equiv \Gamma ' (z)/\Gamma (z)$ denotes the digamma function.
For more details on the analytic properties of the scattering amplitude and related topics
the reader is referred to the review article \cite{Badalian:1981xj}.

We are now in the position to clarify the meaning of the
LETs. Assuming that the functions $F_l
(k^2)$ and $F_l^M (k^2)$ do not possess discrete poles in their
meromorphic regions, the size of the
coefficients in the ERE and MERE (except for the scattering length)
is expected to be governed by the scales $M_L$ and $M_S$ associated with the lowest left-hand singularities,
see \cite{Steele:1998zc} for a related discussion. If the long-range
interaction is known, the quantities $f_l^L (k)$,  $R_l^L (k)$ and $\delta_l^L (k)$
entering the right-hand side of Eq.~(\ref{mere}) can be calculated
explicitly.
Using the MERE, truncated at a given order to approximate
the function $F_l^M (k^2)$, and inverting  Eq.~(\ref{mere}) then allows one
to express the phase shift $\delta_l (k)$ in terms of the known
quantities $f_l^L (k)$,  $R_l^L (k)$ and $\delta_l^L (k)$ and the
first coefficients in the MERE which parametrize physics associated
with the short-range interaction. For example, at LO, the
modified effective range function is approximated
as $F_l^M (k^2) \simeq - 1/a^M$. Thus, using a single piece of
information about the short-range interaction in form of $a^M$ or,
equivalently, the usual scattering length\footnote{Note that   $a^M$ is directly related to  the  scattering length $a$  via
Eq. \eqref{mere}.  Similarly,  the  modified effective range  $r^M$  can be  calculated  using  $a$ and $r$ as input. }  $a$ allows one to predict
\emph{all} coefficients in the ERE. These predictions are accurate up
to corrections emerging from the second term in the MERE.
The MERE for $F_l^M
(k^2)$ thus provides a systematically improvable expansion of the subthreshold parameters entering Eq.~(\ref{ere})
in powers of $M_L/M_S$. The resulting correlations between the
subthreshold parameters at each fixed order in this expansion are referred to as LETs. For an example of an
exactly solvable model, which allows one to get analytical insights into the
meaning of the LETs, the reader is referred to Ref.~\cite{Epelbaum:2009sd}.

We are now in the position to address the LETs for S-wave
neutron-proton scattering. Here and in what follows, we restrict
ourselves to the case of exact isospin symmetry and neglect the
long-range electromagnetic interactions between the neutron and
proton.
The longest-range part of the nuclear force is due to the
OPEP which in momentum space has the form
\beq
\label{OPEPmom}
V_{1\pi} (\vec q \, ) = - \frac{g_A^2}{4 F_\pi^2} \, \frac{\vec \sigma_1 \cdot \vec
q \; \vec \sigma_2 \cdot \vec q}{\vec q \, ^2 + M_\pi^2} \, \fet
\tau_1 \cdot \fet \tau_2 \,,
\eeq
where $\vec q \equiv \vec p \, ' - \vec p$ is the momentum transfer of
the nucleon while $\vec \sigma_i$ ($\fet \tau_i$) denote the spin
(isospin) Pauli matrices of the nucleon $i$. Further, $g_A$ and
$F_\pi$ refer to the axial vector coupling of the nucleon and pion
decay constant, respectively. Throughout this work, we adopt the
values of $F_\pi = 92.4\;$MeV, $g_A=1.267$ for these constants
at the physical value of the pion mass $M_\pi^{\rm phys} \equiv (2
M_{\pi^\pm} + M_{\pi^0} )/3= 138.03\;$MeV.
Upon performing Fourier transform, one
obtains the well-known coordinate-space expression for the long-range
part of the OPEP, namely
\beq
\label{OPEPr}
V_{1\pi} (\vec r \, ) = \frac{M_\pi^3}{12 \pi} \, \frac{g_A^2}{4
  F_\pi^2} \, \fet \tau_1 \cdot \fet \tau_2 \,
\bigg[ S_{12}  \bigg(1 + \frac{3}{M_\pi r}
+ \frac{3}{(M_\pi r)^2} \bigg)
+ \vec \sigma_1 \cdot \vec \sigma_2 \bigg] \, \frac{e^{-M_\pi
    r}}{M_\pi r}\,,
\eeq
where $S_{12}$ is the usual tensor operator, $S_{12} \equiv 3 (\vec
\sigma_1 \cdot \vec r \, )( \vec \sigma_2 \cdot \vec r \, )/r^2 - \vec
\sigma_1 \cdot \vec \sigma_2$. Notice that the above potential
describing point-like nucleons interacting via one-pion exchange
is singular and thus cannot be directly employed in the calculation of
the LETs using the quantum mechanical framework outlined
above.\footnote{It should,
however, be understood that the expressions for the OPEP
in Eqs.~(\ref{OPEPmom}), (\ref{OPEPr}) become meaningless at short
distances where effects due to finite nucleon size cannot be
neglected, see Ref.~\cite{Epelbaum:2006pt} for a related
discussion. Furthermore, while the OPEP clearly dominates the
interaction between the nucleons at large distances, the two-pion exchange
potential becomes comparable in size to the OPEP at distances smaller
than $r \sim 2\;$fm.}  One possible way to overcome this complication
is to apply a properly regularized version of the OPEP. For example,
one can employ a coordinate-space regularization procedure introduced in Refs.~\cite{Gezerlis:2013ipa,Gezerlis:2014zia}
or a closely related method used in
Refs.~\cite{Epelbaum:2014efa,Epelbaum:2014sza}. These approaches are analogous to the Pauli-Villars
regularization and maintain, per construction, the analytic structure
of the amplitude in the sense that non-analyticities   induced by
the regulator appear at momenta $k \geq \Lambda$. Here, $\Lambda$
refers to the momentum scale associated with the regulator which in such an approach should be chosen of the
order of the breakdown scale in the problem
\cite{Lepage:1997cs,Epelbaum:2009sd}. The properly regularized
OPEP can then be used to compute the Jost solution by solving the
Schr\"odinger equation and to work out the LETs for NN scattering.

Instead of using the  approach described above, we employ
here the modified version of Weinberg's chiral EFT proposed in
Ref.~\cite{Epelbaum:2012ua} in order to work out the LETs for NN scattering.  This
momentum-space framework is more convenient for performing
numerical calculations of the LETs and, as already mentioned in the
introduction, has an advantage of being well suited for carrying out chiral extrapolations
beyond the LETs \cite{NNEFT}.

Within the modified Weinberg EFT approach of Ref.~\cite{Epelbaum:2012ua}, the LO amplitude
is obtained by solving the Kadyshevsky equation
\cite{Kadyshevsky:1967rs} which, for the case of the fully off-shell
kinematics, takes the form
\begin{equation}
T\left(p_0,
\vec p\,',\vec p \, \right)=V \left(
\vec p\,',\vec p \, \right) + \int d^3 q \;
 V  \big(
\vec p\,',\vec q \, \big) \ G(p_0,q) \ T  \big(p_0,
\vec q,\vec p\big)
\,,\label{MeqLOk0integrated}
\end{equation}
where $G(p_0,q)$ is the free Green function
\begin{equation}
G(p_0,q)=\frac{m_N^2}{2(2\,\pi)^3}\frac{1}{\big(\vec q\:^2+m_N^2\big)\left(
p_0-\sqrt{\vec q\:^2+m_N^2}+i \epsilon\right)}\,.
\label{G}
\end{equation}
Further, $\vec p$ ($\vec p \, '$) is the incoming (outgoing) three-momentum of
the nucleon in the center-of-mass frame, $p_0 = \sqrt{\vec k  ^2 +
  m_N^2}$ with $m_N$ denoting the nucleon mass and $\vec k$ being the
corresponding three-momentum of an incoming (on-mass-shell) nucleon.
The integral equation (\ref{MeqLOk0integrated}) is solved numerically in the
partial wave basis.

In this work, we are interested in
the $^1$S$_0$ and the coupled $^3$S$_1$-$^3$D$_1$ channels of
neutron-proton scattering. In both cases, the LO potential is given by
the OPEP specified in Eq.~(\ref{OPEPmom}) accompanied by a
momentum-independent contact interaction, whose strength is adjusted
to reproduce the S-wave scattering length. We regularize the integral
equation by introducing an ultraviolet momentum cutoff $\Lambda$ and
take the limit $\Lambda \to \infty$ when calculating the phase
shifts. Notice that contrary to the non-relativistic framework based
on the Lippmann-Schwinger equation, this is a legitimate procedure
since the LO integral equation is renormalizable in the sense that all
divergences emerging from its iterations are absorbable into
redefinition of the LO contact interaction. Once the phase shifts are
calculated, we extract the effective range and the shape parameters
numerically by matching the truncated ERE to the
effective range function. Given that the calculated S-wave scattering amplitudes
fulfill elastic unitarity, correctly reproduce the first left-hand cut due to
the OPEP and match the experimental value at the threshold (given by
the scattering length), the resulting predictions for $r$ and $v_i$ are equivalent to
the LO LETs discussed above in the context of the MERE. We emphasize,
however, that employing the cutoff-independent framework of
Ref.~\cite{Epelbaum:2012ua} is not crucial for testing the LETs in NN
scattering. A more conventional non-relativistic chiral EFT approach
based on the Lippmann-Schwinger equation and utilizing a finite
ultraviolet cutoff is equally well suited for this purpose.

\begin{figure}[tb]
\vskip 1 true cm
\includegraphics[width=0.8\textwidth,keepaspectratio,angle=0,clip]{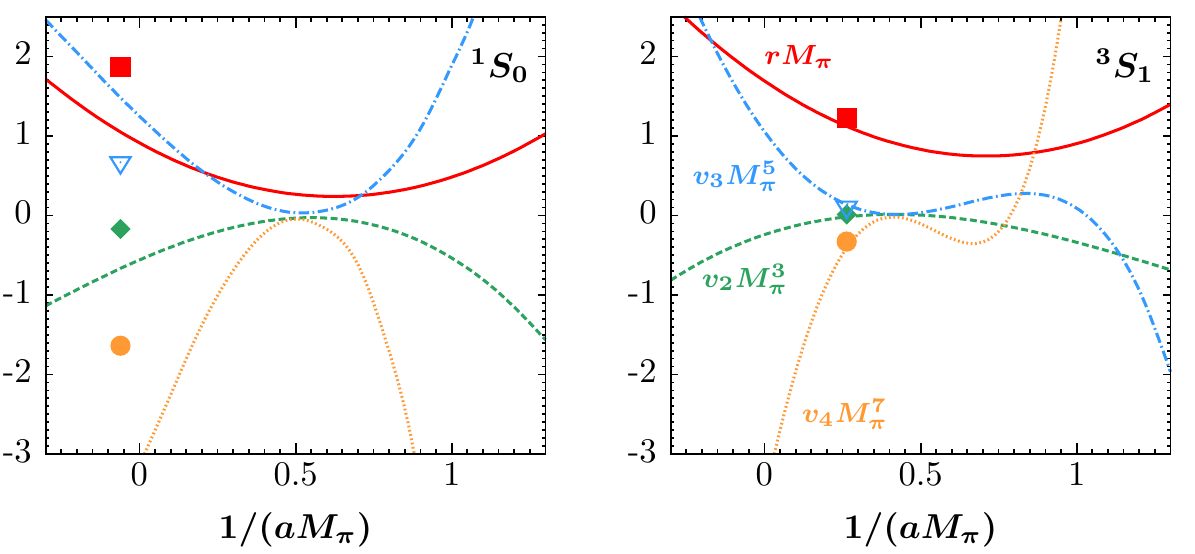}
    \caption{Correlations between the inverse scattering length $a^{-1}$, effective range $r$ and the first
three shape parameters $v_2$, $v_3$ and $v_4$ induced by the one-pion exchange interaction in the  $^1$S$_0$
(left panel) and $^3$S$_1$ (right panel) channels. Solid rectangles, diamonds, open triangles and circles  correspond to the values
of $r$, $v_2$, $v_3$ and $v_4$, respectively,  extracted from the Nijmegen partial
wave analysis \cite{Stoks:1994wp,PavonValderrama:2005ku}.
The results in the $^3$S$_1$
partial wave correspond to the Blatt-Biedenharn parametrization of the
S-matrix \cite{Blatt:1952zz}.
\label{fig:1S0_3S1}
 }
\end{figure}
The LO LETs for the $^1$S$_0$ and the $^3$S$_1$ channels have
already been addressed in Ref.~\cite{Epelbaum:2012ua}. In
Fig.~\ref{fig:1S0_3S1}, we show the LO LET predictions for the effective range and
the first three shape parameters as functions of the inverse
scattering length which is an input parameter in our
calculations. Here and in what follows, the results in the $^3$S$_1$
partial wave correspond to the Blatt-Biedenharn parametrization of the
S-matrix \cite{Blatt:1952zz}.
To avoid the appearance of large/small numbers, we use
dimensionless quantities by multiplying the coefficients in the ERE
with the corresponding powers of the pion mass. As expected,
the predicted dimensionless coefficients appear to be
of a natural size for a rather large range of values of the
scattering length. Notice further that the limit $a \to \pm \infty$,
which
describes the
situation in which there is a zero-energy bound state, does not
lead to any peculiarities in $r$ and $v_i$. In both channels, one
observes a clear tendency for $r$ and $v_i$ to become
unnaturally large in magnitude in the regions of  $1/(a M_\pi) \lesssim -0.5$
and  $1/(a M_\pi) \gtrsim 1$.  We found that the unnaturally large values
of $r$ and $v_i$ for $1/(a M_\pi) \gtrsim 1$ are due  to the
appearance of a pole in the effective range function at relatively
low positive energy. The pole position corresponds to the energy
at which the corresponding phase shift crosses zero. For example, for $1/(a M_\pi) \simeq 1.4$, the
pole in the $^3$S$_1$ effective range function is located at $k \simeq
60$ MeV which is  already within its meromorphic region. For larger values of
$1/(a M_\pi)$, the pole moves towards the threshold so that the values of
the $r$ and $v_i$ are actually governed by the pole position rather than
by $M_\pi$. As already emphasized above, the appearance of a pole in
the effective range function can be easily accommodated
by replacing the Taylor expansion of $F_l (k^2)$
by e.g.~Pade approximation.
On the other hand,  the
case $1/(a M_\pi) \lesssim -0.5$ corresponds to the absence of
a bound state and,  given the  natural value of the scattering length,  it might already
describe a perturbative regime of NN scattering. For weakly
interacting systems, it is easy to see analytically (by making use of the Born
approximation) that the coefficients in the ERE do not scale anymore
with the range of the interaction. This is due to the appearance of an
additional dimensionless small parameter associated with the weakness
of the interaction.

The predictions for $r$ and $v_i$ based on the  LO LETs and
corresponding to the experimentally observed values of the scattering
lengths are compared in Fig.~\ref{fig:1S0_3S1}
with the empirical numbers from Ref.~\cite{Stoks:1994wp,PavonValderrama:2005ku}
based on the Nijmegen partial wave analysis
(PWA).  We also collect various available results for the LETs in the  $^1$S$_0$ and the $^3$S$_1$
channels in Table \ref{LET_3S1_physical}.
\begin{table}[t]
\caption{Various available results for the low-energy theorems for the
  neutron-proton $^1$S$_0$ and $^3$S$_1$ partial
  waves as described in the text. The results in the $^3$S$_1$
partial wave correspond to the Blatt-Biedenharn parametrization of the
S-matrix \cite{Blatt:1952zz}.
\label{LET_3S1_physical}}
\smallskip
\begin{tabular*}{\textwidth}{@{\extracolsep{\fill}}lrrrrr}
\hline
\hline
\noalign{\smallskip}
  &   $a$ [fm] &     $r$ [fm] &  $v_2$ [fm$^3$] &  $v_3$ [fm$^5$] &  $v_4$ [fm$^7$]
\smallskip
 \\
\hline
\hline
\multicolumn{6}{l}{\bf \hskip -0.1 true cm Neutron-proton $^1$S$_0$ partial wave } \\
LO, Ref.~\cite{Epelbaum:2012ua}  &  fit  & $1.50$  & $-1.9$ &$8.6(8)$  &$-37(10)$
\\[2pt]
NLO, nonperturbative $C_2$, Ref.~\cite{Epelbaum:2015sha}  &   fit  & fit  &
$-0.61 \ldots -0.55$ &$ 5.1 \ldots 5.5 $  &$-30.8 \ldots -29.6$
\\ [2pt]
\hline
NLO KSW, Ref.~\cite{Cohen:1998jr}  & fit   & fit  & $-3.3$  & $18$ & $-108$
\\ [2pt]
&&&&& \\ [-12pt]
\hline
Empirical values, Ref.~\cite{PavonValderrama:2005ku} & $-23.7$  & $2.67$  & $-0.5$  & $4.0$ & $-20$
\\ [2pt]
\hline \hline
&&&&& \\ [-12pt]
\multicolumn{6}{l}{\bf \hskip -0.1 true cm Neutron-proton $^3$S$_1$ partial wave } \\
LO, Ref.~\cite{Epelbaum:2012ua}  & fit & $1.60$ & $-0.05$ & $0.82$ & $-5.0$  \\  [2pt]
NLO, this work  & fit & fit & $0.06$ & $0.70$ & $-4.0$  \\ [2pt]
\hline
NLO KSW, Ref.~\cite{Cohen:1998jr}  & fit   & fit  & $-0.95$  & $4.6$ &
                                                                       $-25$
\\ [2pt]
&&&&& \\ [-12pt]
\hline
Empirical values, Ref.~\cite{deSwart:1995ui}  &  $5.42$ & $1.75$ & $0.04$ & $0.67$ & $-4.0$ \\
[4pt]
%\smallskip
\hline
\hline
\end{tabular*}
\end{table}
The results for the LO LETs shown in the table are taken from
Ref.~\cite{Epelbaum:2012ua}. While the predicted values of
$r$ and $v_i$ do  reproduce
the empirical values in the $^3$S$_1$ channel rather accurately,  the agreement in the
$^1$S$_0$ partial wave is, at best, on a  qualitative level. This pattern
can  be  understood as follows. Generally, the accuracy of the LO LET
is expected to be set by the ratio $E_L/E_S$, where $E_L$ ($E_S$)
denotes the energy corresponding to the branch point of the left-hand
cut due to the long-range interaction (branch point of the first left-hand
cut in the scattering amplitude which is not correctly described by
the employed approximation for the interaction). Therefore, given that we
do not take into account the two-pion exchange potential in our
calculations and thus cannot describe correctly the second left-hand
cut,
one may expect the LO LETs to be accurate at the level
of $\sim 25\%$. In fact, the accuracy of the LO LETs in the $^3$S$_1$ partial wave
appears even to be somewhat higher.\footnote{The large relative
  deviation for the first shape parameter should not be taken too
  seriously due to its unnaturally small value.} This is presumably due to the
fact, that we actually include the parts of the second and higher left
hand cuts (see Fig.~\ref{fig:cuts}) which are associated with iterations of the OPEP. In
fact, according to the chiral power counting, such iterative
contributions to the scattering amplitude are expected to be more
important than the irreducible two- and more-pion exchange potentials.
On the other hand, the low accuracy of the LETs in the $^1$S$_0$ channel is
a consequence of the weakness of the OPEP projected on that partial
wave. The strong tensor part of the OPEP, which accompanies the operator
$S_{12}$ in Eq.~(\ref{OPEPr}), does not contribute to spin-singlet partial
waves. The weakness of the OPEP in the $^1$S$_0$ channel is evidenced
by the fact, that the phase shift corresponding to the
long-range OPEP  reaches only about $12.6 ^\circ$ at maximum, which
has to be compared with $max (\delta_{1S0} (k) ) \sim 65^\circ$ in the
real world.

We also list in Table
\ref{LET_3S1_physical}, for the sake of completeness, the predictions obtained
within the KSW framework
\cite{Kaplan:1998we}  which relies upon a perturbative treatment of the OPEP.
These results are taken from Ref.~\cite{Cohen:1998jr} and
show very large deviations from the empirical values. This shows
that the
perturbative treatment of the OPEP in these channels is
not appropriate.

In Ref.~\cite{Epelbaum:2015sha}, the subleading contact
interaction $C_2 (p^2 + p'^2)$ has been taken into account in the
$^1$S$_0$ channel within the framework of Ref.~\cite{Epelbaum:2012ua}, both perturbatively and nonperturbatively. Given that the
long-range interaction included in that work is entirely given by the
OPEP, the resulting predictions for the shape parameters, which are
listed in Table \ref{LET_3S1_physical} for the case of the
nonperturbative treatment of $C_2$, are equivalent to the
next-to-leading-order (NLO) LETs.
Notice that the non-perturbative inclusion of the subleading contact
interaction necessarily results in a residual subtraction scale
dependence of the amplitude which is why  the predictions for $v_{2,3,4} $ are shown within a range of  values,
 see Ref.~\cite{Epelbaum:2015sha} for more details.
While  a significant improvement is observed in the description
of all $v_i$, the convergence is clearly rather slow in this channel.
We find that the accuracy of the LETs in the $^1$S$_0$ channel
is insufficient for the purpose of providing constraints on the available
lattice QCD results. For this reason,  we will concentrate in the following entirely on the
spin-triplet channel.

While the LETs  in the $^3$S$_1$ channel appear to be fairly accurate already
at LO as far as one is concerned with predicting the values of $r$ and
$v_i$ from the scattering length, their accuracy is still insufficient
for extracting the deuteron binding energy using the effective range as  input, as we
intend to do in the next sections. This is
because of the function $r(a^{-1})$ being rather flat in the relevant region of values
of $1/(a M_\pi)$, which tends to magnify the uncertainty in the
value of $a^{-1}$ extracted from $r$.  Notice that  the observed nearly  quadratic dependence of  $r$  on  $a^{-1}$  may be  related  to the universal behavior
of this function found for van der Waals-like  interactions  \cite{PavonValderrama:2005wv}, see also Ref.~\cite{Elhatisari:2013swa} for a related discussion.
In addition, the deuteron binding energy
depends, to a good accuracy, quadratically on $a^{-1}$, so that
the error in the value of the binding energy  is further
magnified. For example, using the LO LETs together with the  experimentally
observed value of the effective range $r = 1.75\;$fm as  input, one
extracts $a=7.2\;$fm and  $B_d = 1.1\;$MeV which is two
times smaller than the experimentally observed value of $B_d \simeq
2.22\;$MeV. To deal with this issue, we extend the LETs in the
$^3$S$_1$ channel to NLO by taking into account a subleading
short-range interaction whose strength is adjusted to reproduce the
experimentally observed value of the effective range. Notice that
contrary to the $^1$S$_0$ channel, where the long-range part of the
OPEP is non-singular and a non-perturbative inclusion of the
subleading contact interaction can be carried out semi-analytically in
a close analogy to pionless EFT, see Ref.~\cite{Epelbaum:2015sha} for
more details, we are unable to treat the subleading contact
interactions non-perturbatively in spin-triplet channels without, at the
same time, destroying the explicit renormalizability feature of the scattering amplitude.
A perturbative inclusion of the NLO short-range terms will be discussed
in detail in a  subsequent publication  \cite{NNEFT}.
Here, in order to avoid possible
shortcomings due to  the perturbative treatment of the
subleading short-range terms, we follow a different approach and
employ resonance saturation to model higher-order contact interactions
by means  of a heavy-meson-exchange. In particular, we use
for the NLO potential
\beq
\label{satur}
V_{\rm NLO} = V_{1 \pi} (\vec q \, ) + C_0 + \beta \, \frac{\vec \sigma_1 \cdot \vec
q \; \vec \sigma_2 \cdot \vec q}{\vec q \, ^2 + M^2}\,,
\eeq
where the heavy-meson mass $M$ is set $M=700$ MeV and the strength
$\beta$ is adjusted to reproduce the empirical value of the effective
range.  In agreement with the arguments given  above,  we have verified that  the results  for  the LETs are insensitive
to   details  of  the short-range interaction.
In particular, the  deviation in the results caused by using  the  sigma-like  scalar potential  instead of the tensor one in Eq.~(\ref{satur})
 for the sub-leading short-range interaction is negligibly small.
The sub-leading short-range interaction  is only needed to  describe   the second term in the MERE   $\sim r^M$
which appears, in addition to $a^M$,  as an input parameter in our calculation.
The employed form of the NLO potential ensures that the corresponding
scattering amplitude is renormalizable so that the NLO calculations can be
carried out in the same way as at LO. The predicted values of the
shape parameters after fixing the values of  $C_0$ and $\beta$ are listed in Table~\ref{LET_3S1_physical} and, as
expected, show a clear improvement as compared with the LO results. In
fact, the NLO LETs appear to be accurate at the level of a few
percent (except for $v_2$).

%%%%%%%%%%%%%%%%%%%%%%%%%%%%%%%%%%%%%%%%%%%%%%%%%%%%%%%%%%%%%%%%%%%%%%%%%%%%%%%%%
\section{Low-energy theorems for unphysical pion masses}
\def\theequation{\arabic{section}.\arabic{equation}}
\label{sec:LET}

\begin{figure}[tb]
\vskip 1 true cm
\includegraphics[width=\textwidth,keepaspectratio,angle=0,clip]{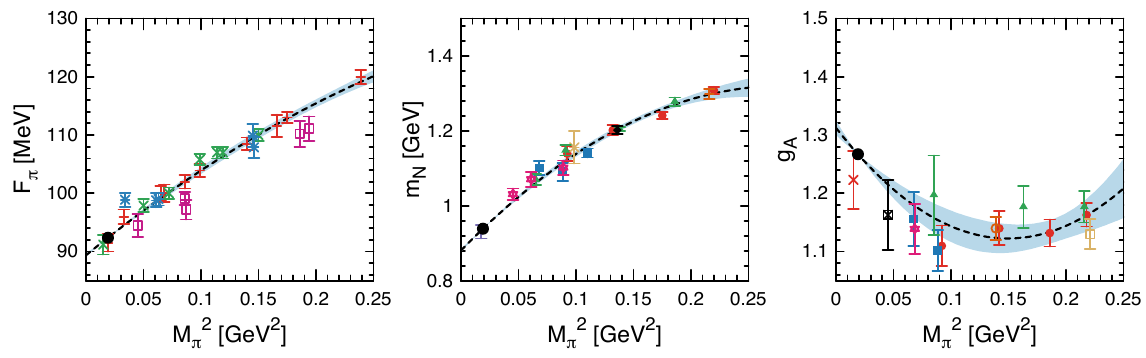}
    \caption{Quadratic polynomial regression fits to lattice QCD data for the pion decay constant $F_\pi$,
nucleon mass $m_N$ and the nucleon axial-vector
coupling constant $g_A$. Lattice data for $F_\pi$ and  $m_N$ ($g_A$) correspond to the $2+1$ flavor simulations
of the Budapest-Marseille-Wuppertal collaboration using Wilson fermions \cite{Durr:2013goa} and $2$ and $2+1+1$ flavor
simulations reported in Ref.~\cite{Alexandrou:2013jsa} (Ref.~\cite{Alexandrou:2014wca}) using twisted mass fermion ensembles.
For details of the simulations and analyses see the above mentioned references.
Filled circles without error bars show the experimentally measured values. The dashed lines depict our fit results to the lattice
data  and the experimental values while the shaded bands correspond to the $67\%$ confidence levels of the interpolations.
\label{fig:Input}
 }
\end{figure}

The LETs discussed in the previous section can be straightforwardly
generalized to the case of non-physical quark (or pion) masses.  The
main effect from changing the pion mass, as far as the analytic
structure of the amplitude is concerned,  corresponds to
shifts of  the branch points of the left-hand cuts due to one-, two- and
more-pion exchange, see Fig.~\ref{fig:cuts}. This effect originates from
the explicit pion mass dependence in the pion propagator. Next, one
needs to account for the change in the discontinuity across the
left-hand cuts caused by the $M_\pi$ dependence of the strength of the
OPEP, namely the ratio $g_A/F_\pi$. To account for this effect, we
make use of the  lattice-QCD results for the
$M_\pi$ dependence of these quantities, see Fig.~\ref{fig:Input}.
In particular, we performed quadratic polynomial regression fits (as functions of $M_\pi^2$) of the
results from the $2+1$ flavor simulations
by the Budapest-Marseille-Wuppertal (BMW) collaboration for $F_\pi$
\cite{Durr:2013goa}  and from the $2$ and $2+1+1$ flavor
simulations reported in Ref.~\cite{Alexandrou:2014wca} for
$g_A$, see also Ref.~\cite{Horsley:2013ayv} for recent lattice-QCD calculations of
this quantity. In our fits, we included
lattice-QCD data for pion masses up to $M_\pi = 500\;$MeV. Further,
we enforced the fits to go through the experimental value for a given
quantity at the physical value of the pion mass.
The results of our fits together with the $67\%$ confidence region are
shown  in Fig.~\ref{fig:Input}. Notice that the observed
tendency of lattice-QCD results for the axial charge of the nucleon to
somewhat underestimate the experimental value may indicate that
certain systematic corrections have not been properly taken into
account in these studies. We expect this issue to be clarified in the
near future, when more results near the physical point will become
available and the statistical uncertainties of the simulations will be
improved. Notice further that we did not include all available
lattice QCD results in our fits since this will unlikely improve the
accuracy of the  interpolations. Especially for the pion decay
constant, the uncertainty of the  interpolation of the
lattice-QCD results is very small and can be safely neglected at the
accuracy level of our calculations.

Next, one also has to account for the pion mass dependence of the
nucleon mass. Here, we follow the same strategy as for $F_\pi$ and $g_A$ and employ a quadratic
interpolation of the recent lattice-QCD results for $m_N(M_\pi^2)$ of the BMW collaboration
\cite{Durr:2013goa}, see Fig.~\ref{fig:Input}. We emphasize that our
fits of the $M_\pi$ dependence of $F_\pi$, $g_A$ and $m_N$ are not
intended to provide correct chiral extrapolations of these
quantities to the chiral limit but serve exclusively to enable a
smooth interpolation of the available lattice-QCD results.

\begin{figure}[tb]
\vskip 1 true cm
\includegraphics[width=1.0\textwidth,keepaspectratio,angle=0,clip]{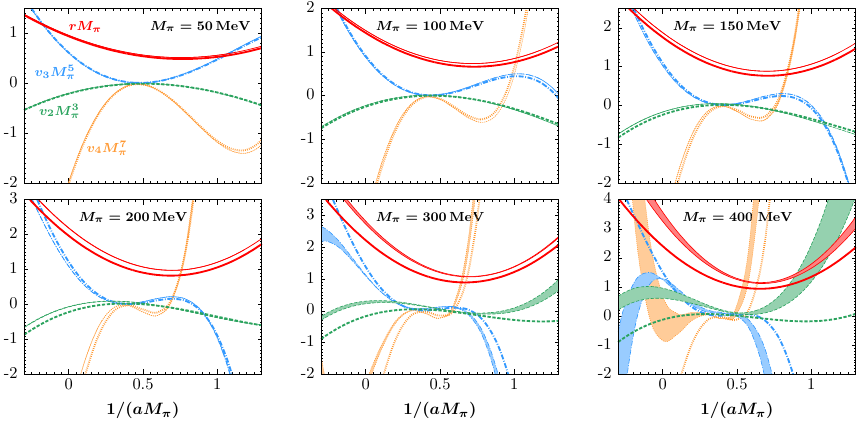}
    \caption{Correlations between the inverse scattering length $a^{-1}$, effective range $r$ and the first
three shape parameters $v_2$, $v_3$ and $v_4$ in the $^3$S$_1$   partial wave induced by the one-pion
exchange interaction. Various thick lines show the predictions of the LO LETs while light-shaded bands (hardly visible for small $M_\pi$) between thin lines depict
the results of NLO LETs and reflect the estimated uncertainty due to
unknown $M_\pi$ dependence
of the subleading short-range interaction as explained in the text.
\label{fig:LET_NLO}
 }
\end{figure}

Using the above results
for the pion mass dependence of $F_\pi$, $g_A$ and $m_N$ together with
the explicit $M_\pi$ dependence in the pion propagator allows us to describe the
discontinuity across the first left-hand cuts due to the OPEP for
arbitrary values of the pion masses. Thus, we can immediately generalize
the LO LETs by calculating $r$ and $v_i$ as functions of the (inverse)
scattering length at unphysical values of the pion mass.  The resulting
predictions are shown by the various thick lines in Fig.~\ref{fig:LET_NLO} for
$M_\pi=50$, $100$, $150$, $200$, $300$ and $400\;$MeV.   Notice that  contrary  to chiral EFT  extrapolations,
we do not make here any assumptions about the quark mass dependence of the
included derivative-less short-range interaction $\propto C_0$, which is traded into
the dependence of the calculated observables on the value of the scattering length.
In particular,  we  perform calculations  with different  pion masses as if we lived in  different worlds, in which the physical value of the pion mass were
$M_\pi=50$, $100$,   $\dots \;$MeV.
Thus  for each given pion mass the value
of  $C_0$ is adjusted to reproduce the  given value of the  scattering length used as input.
In this sense, our results are more
general than the ones obtained for the $^1$S$_0$ and $^3$S$_1$
channels in Ref.~\cite{Epelbaum:2013ij} at LO within the modified Weinberg approach,
where the dependence of the short-range interaction on $M_\pi$ was
neglected in agreement with the estimation based on naive dimensional
analysis (NDA).

In order to extend these results to NLO, we need to specify the
$M_\pi$ dependence of the subleading short-range interaction which
is modeled via resonance saturation, see Eq.~(\ref{satur}). While
this effect would be suppressed in the chiral EFT approach if one assumes the
validity of NDA for short-range interactions, we decided to estimate
it in order to stay as model independent as
possible. Specifically, we assume that the $M_\pi^2$ dependence of the
strength $\beta$ of the short-range interaction is within the envelope
built by the lines which go through the physical point and describe a
$\pm 50\%$ change in the value of $\beta$ for $M_\pi = 500$
MeV, i.e.:
\begin{equation}
\label{beta_range}
1 - \delta \beta \left|\frac{M_\pi^2 -(M_\pi^{\rm phys})^2}{\Delta
    M_\pi^2} \right|    \leq \frac{\beta (M_\pi)}{\beta (M_\pi^{\rm
    phys})} \leq 1 + \delta \beta \left|\frac{M_\pi^2 -(M_\pi^{\rm phys})^2}{\Delta
    M_\pi^2} \right|\,,
\end{equation}
with $\delta \beta = 0.5$ and $\Delta M_\pi^2 \equiv (M_\pi^2 -
(M_\pi^{\rm phys})^2 )\big|_{M_\pi = 500\; \mbox{MeV}}$.
Note that such an estimation is justified for the considered
quantities such as $g_A$, $F_\pi$ and $m_N$, as  can be seen  from Fig. \ref{fig:Input}.

Our NLO LETs predictions for the effective range and the shape parameters
viewed as functions of the inverse scattering length are visualized in
Fig.~\ref{fig:LET_NLO} by the light-shaded bands. These bands correspond to the
variation of $\beta$ at a given value of $M_\pi$ according to
Eq.~(\ref{beta_range}). For pion masses below $M_\pi \simeq
200\;$MeV, one observes a rather small difference between the
predictions based on LO and NLO LETs whose size can be viewed as an
estimation of the accuracy of the LO LETs. Also, the
employed variation of $\beta$ is essentially invisible for such values
of the pion mass. For heavier pions, both the differences between the LO
and NLO LETs as well as the uncertainty in the results associated with
the $M_\pi$ dependence of $\beta$ start to increase. As shown in this
figure, one cannot expect to have accurate predictions for pion masses above
$M_\pi \simeq 400\;$MeV. Notice that decrease in the
predictive power of the LETs is to be expected for heavier pions due
to the decreasing separation between the soft and hard scales in the
problem.

%%%%%%%%%%%%%%%%%%%%%%%%%%%%%%%%%%%%%%%%%%%%%%%%%%%%%%%%%%%%%%%%%%%%%%%%%%%%%%%%%
\section{Implications for lattice QCD calculations}
\def\theequation{\arabic{section}.\arabic{equation}}
\label{sec:LatticeQCD}

\begin{figure}[tb]
\vskip 1 true cm
\includegraphics[width=0.8\textwidth,keepaspectratio,angle=0,clip]{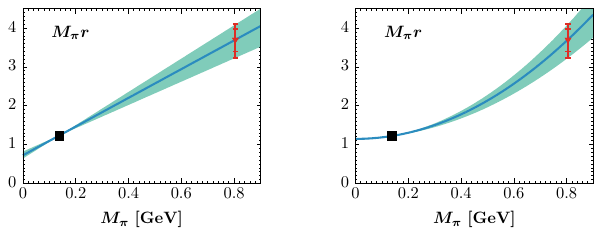}
    \caption{Left panel: Linear with $M_\pi$ interpolation of the
      quantity $M_\pi r$  in the $^3$S$_1$ partial wave according to
      Eq.~(\ref{r_extrap}) as suggested in Ref.~\cite{Beane:2013br}. Right
      panel: Linear with $M_\pi^2$ interpolation of the
      quantity $M_\pi r$  in the $^3$S$_1$ partial wave according to
      Eq.~(\ref{r_extrap2}). In both cases, solid squares refer to the value of $M_\pi r$ at the physical point.
\label{fig:RangeInput}
 }
\end{figure}

We are now in the position to confront the LETs with the
available lattice QCD results in the NN sector. In
table \ref{data}, we list the published lattice-QCD results for the
S-wave scattering parameters and energies of the bound states together
with the experimental data. We do not show in the table the results
from Ref.~\cite{Beane:2006mx} where volume dependence was not
addressed.
\begin{table}[t]
\caption{Available experimental and infinite-volume lattice QCD data for nucleon-nucleon scattering parameters and
bound state energies in the $^1$S$_0$ and $^3$S$_1$ channels at various values of the pion mass.
\label{data}}
\smallskip
\begin{tabular*}{\textwidth}{@{\extracolsep{\fill}}rrrrrr}
\hline
\hline
\noalign{\smallskip}
  &  $M_\pi = 138$ MeV &   $M_\pi = 300$ MeV \cite{Yamazaki:2015asa}&     $M_\pi = 390$ MeV \cite{Beane:2011iw} &  $M_\pi=510$ MeV \cite{Yamazaki:2012hi} & $M_\pi = 800$ MeV \cite{Beane:2013br}
\smallskip
 \\
\hline
\hline
\multicolumn{6}{l}{The $^3$S$_1$ channel } \\
$B_d$ [MeV] & $2.224$ &  $14.5(0.7)({}^{+2.4}_{-0.7})$  & $11(05)(12)$ &  $11.5(1.1)(0.6)$ & $19.5(3.6)(3.1)(0.2)$\\  [2.5pt]
$a$ [fm] & $5.42$ & not given  & not given  & not given & $1.82({}^{+0.14}_{-0.13})({}^{+0.17}_{-0.12})$ \\  [2.5pt]
$r$ [fm] & $1.75$ &  not given & not given & not given  &  $0.906({}^{+0.068}_{-0.075})({}^{+0.068}_{-0.084})$ \\
[4pt]
\hline
\multicolumn{6}{l}{The $^1$S$_0$ channel } \\
$B_{nn}$ [MeV] & -- &  $8.5(0.7)({}^{+2.2}_{-0.4})$  & $7.1(5.2)(7.3)$ &  $7.4(1.3)(0.6)$ & $15.9(2.7)(2.7)(0.2)$\\   [2.5pt]
$a$ [fm] & $-23.7$ & not given  & not given & not given &  $2.33({}^{+0.19}_{-0.17})({}^{+0.27}_{-0.20})$ \\  [2.5pt]
$r$ [fm] & $2.67$ & not given  & not given & not given   &  $1.130({}^{+0.071}_{-0.077})({}^{+0.059}_{-0.063})$  \\
[4pt]
%\smallskip
\hline
\hline
\end{tabular*}
\end{table}
Unfortunately, lattice calculations in the NN sector  focused so far mainly
on the binding energies and do not provide information on
the scattering parameters. An exception is the work of
Ref.~\cite{Beane:2013br}, which provides, in addition to the binding
energies, also the values of the scattering length, effective range
and even the first shape parameter at the pion mass of $M_\pi \simeq
800\;$MeV. Clearly, such heavy pion masses are out of reach of the LETs.
On the other hand, the authors of Ref.~\cite{Beane:2013br} conjectured that
the quantity $M_\pi r$ may exhibit a nearly linear dependence on
the pion mass. The suggested linear interpolation between
the physical point and the lattice result has the form \cite{Beane:2013br}:
\begin{equation}
\label{r_extrap}
M_\pi r \cong A^{(^3S_1)} + B^{(^3S_1)} M_\pi, \quad \;
\mbox{where}\; \; \; A^{(^3S_1)} =
0.726^{+0.065}_{-0.059}{}^{+0.072}_{-0.059} \,, \; \; \;
B^{(^3S_1)} =
3.70^{+0.42}_{-0.47}{}^{+0.42}_{-0.52} \; \;\mbox{GeV}^{-1}\,,
\end{equation}
and is visualized in the left panel of Fig.~\ref{fig:RangeInput}.
While we cannot judge on the validity of the suggested linear
dependence of the quantity $M_\pi r$ on the pion mass based on the
LETs alone,
we can test its compatibility with the
lattice-QCD results for the deuteron binding energy available for pion
masses within the validity range of the LETs. Specifically, we employ the
effective range $r (M_\pi)$ from  Eq.~(\ref{r_extrap}) instead of the
scattering length to fix the $M_\pi$ dependence of the short-range
interaction by adjusting the value of $C_0$ in Eq.~(\ref{satur}) and
make predictions for the scattering length, shape parameters and the
deuteron binding energy.
\begin{figure}[tb]
\vskip 1 true cm
\includegraphics[width=1.0\textwidth,keepaspectratio,angle=0,clip]{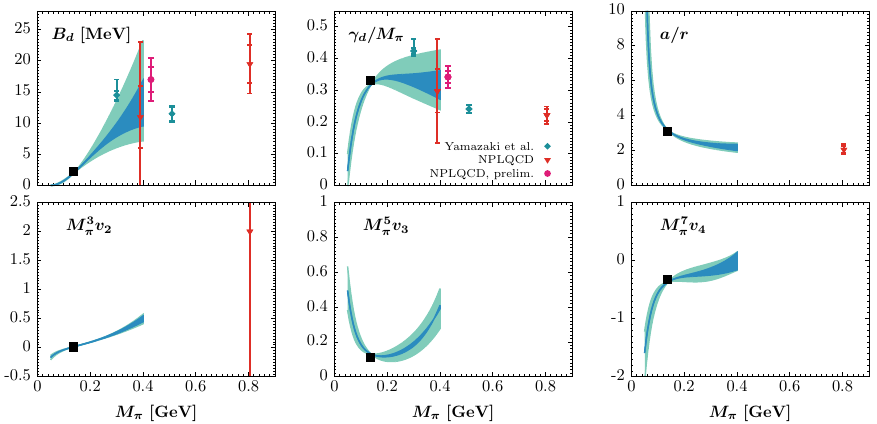}
    \caption{NLO LET predictions for the pion mass dependence of the
deuteron binding energy, the ratio $\gamma_d/M_\pi$, the ratio
$a/r$ and the first three shape parameters in the $^3$S$_1$ partial wave assuming the
linear $M_\pi$ dependence of the effective range specified in
Eq.~(\ref{r_extrap}) and visualized in the left
panel of Fig.~\ref{fig:RangeInput}.
Dark-shaded bands show our estimation of the uncertainty of the NLO
LETs due to the unknown $M_\pi$ dependence
of the subleading short-range interaction specified in Eq.~(\ref{beta_range}), light-shaded bands depict the uncertainty in the
linear extrapolation of the effective range used as input, as shown in
the left panel of Fig.~\ref{fig:RangeInput}.
\label{fig:predictions}
 }
\end{figure}
Our results for the deuteron binding energy $B_d$, the ratio
$\gamma_d/M_\pi$, where $\gamma_d = \sqrt{B_d m_N}$ is the deuteron
binding momentum, the ratio $a/r$ and the first three shape parameters
$M_\pi^3 v_2$, $M_\pi^5v_3$ and $M_\pi^7 v_4$ are visualized in
Fig.~\ref{fig:predictions}. In this figure, the dark shaded bands
result from the variation of the constant $\beta$ specified in Eq.~(\ref{beta_range})  and reflect the
uncertainty of the NLO LETs\footnote{Note that employing the scalar sub-leading potential in Eq.~\eqref{satur} instead of the tensor one  yields the results
which are well within the  dark shaded band for all quantities   except for the parameter $v_3$ which is relatively small and
appears to be slightly outside of this band  for $M_{\pi} > 250$ MeV.}.  The light-shaded bands  correspond to the resulting uncertainty  which
emerges from  the theoretical  uncertainty at NLO and
the errors of  the linear
interpolation of $M_\pi r (M_\pi)$ (see the left panel of
Fig.~\ref{fig:RangeInput} and Eq.~(\ref{r_extrap}))   added in quadrature.
Notice that we also show in
Fig.~\ref{fig:predictions} the preliminary lattice-QCD result of the NPLQCD collaboration
at $M_\pi = 430\;$MeV \cite{Beane:Prelim}. Remarkably, the
linear $M_\pi$ dependence of $M_\pi r$ suggested in
Ref.~\cite{Beane:2013br} indeed appears to describe very well the
common trend of the lattice-QCD results for the deuteron binding
energy at intermediate pion masses. Also the NPLQCD Collaboration results of Ref.~\cite{Beane:2013br}
for  $B_d$, $\gamma_d/M_\pi$, $a/r$ and $M_\pi^3 v_2$ at the
pion mass of $M_\pi = 800\;$MeV can be well described by
further extrapolating our results to heavier pion masses without introducing any strong
curvature. Assuming the validity of Eq.~(\ref{r_extrap}) for pion
masses below the physical one, we conclude that the deuteron becomes
unbound for $M_\pi \sim 50\;$MeV.  It is also interesting to notice that the scattering
length and the shape parameters show rather strong variations with the
pion mass around and below the physical point. This nontrivial
behavior is driven by the long-range physics associated with the
pion exchange and is, in principle, testable in lattice
QCD. The obtained results for the quantities $\gamma_d /M_\pi$ and
$a/r$, which probe the amount of fine tuning in the NN system, suggest
that the physically realized value of the quark mass is close to the
point, which separates the strong fine-tuning regime characterized by
the rapidly growing scattering length from the regime featuring a
fairly small amount of fine tuning with $a/r = 2\ldots 3$ within the
large range of pion masses.

\begin{figure}[tb]
\vskip 1 true cm
\includegraphics[width=1.0\textwidth,keepaspectratio,angle=0,clip]{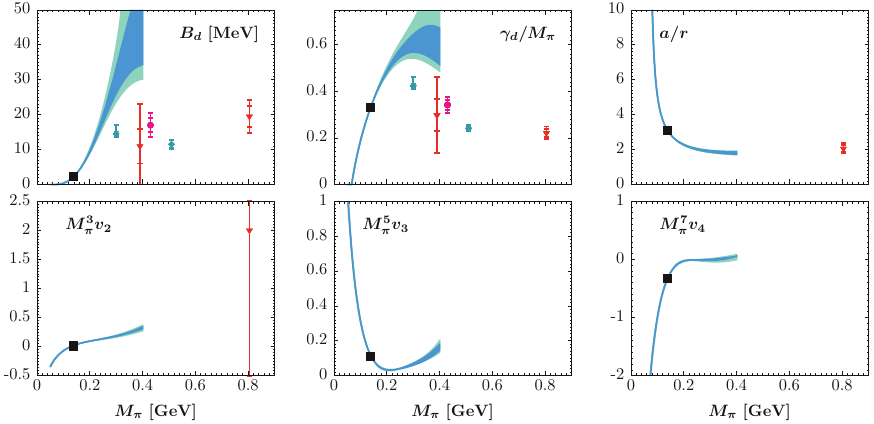}
    \caption{NLO LET predictions for the pion mass dependence of the
deuteron binding energy, the ratio $\gamma_d/M_\pi$, the ratio
$a/r$ and the first three shape parameters in the $^3$S$_1$ partial wave assuming the
linear $M_\pi^2$ dependence of the effective range specified in
Eq.~(\ref{r_extrap2}) and visualized in the right panel of Fig.~\ref{fig:RangeInput}.
For notations see Fig.~\ref{fig:predictions}.
\label{fig:predictions2}
 }
\end{figure}

We emphasize that the observed agreement between the predicted
$M_\pi$ dependence of the deuteron binding energy and lattice-QCD
results is a rather nontrivial consequence of the assumed
extrapolation of $M_\pi r$. To illustrate this point, we consider an
alternative scenario by assuming that $M_\pi r$ is a linear function
of $M_\pi^2$ rather than of $M_\pi$. Notice that while a linear in
$M_\pi$ dependence seems to be
natural for NN observables,
such a scenario cannot be excluded a priori.
Using the lattice-QCD result of
Ref.~\cite{Beane:2013br}  for $M_\pi r$ at the pion mass of  $M_\pi =
800\;$MeV, the resulting interpolation formula takes the form
\begin{equation}
\label{r_extrap2}
M_\pi r \cong C^{(^3S_1)} + D^{(^3S_1)} M_\pi^2, \quad \;
\mbox{where}\; \; \; C^{(^3S_1)} =
1.149^{+0.009}_{-0.009}{}^{+0.011}_{-0.009} \,, \; \; \;
D^{(^3S_1)} =
3.95^{+0.45}_{-0.49}{}^{+0.45}_{-0.55} \; \;\mbox{GeV}^{-2}\,,
\end{equation}
see the right panel of Fig.~\ref{fig:RangeInput}.
The resulting LET predictions for the deuteron binding energy are shown in
Fig.~\ref{fig:predictions2}. The assumed quadratic
dependence on $M_\pi$ is clearly  in conflict with the lattice-QCD results
and, therefore, appears to be highly unlikely.
Interestingly, the shape parameters seem to be more robust with
respect to the variation of the functional form of $r(M_\pi )$ and
show a qualitatively similar behavior in both scenarios.

Finally, as a last application of the LETs, we use the quoted values of the
deuteron binding energy to predict the scattering length, effective
range and the first three shape parameters at pion masses of $M_\pi =
300\;$MeV,  $M_\pi =390\;$MeV and $M_\pi =430\;$MeV where lattice-QCD
results are already available. Our results for the coefficients in the
ERE are collected in Table \ref{predictions}.
\begin{table}[t]
\caption{Predictions for the scattering length, effective range and
  shape parameters based on the lattice-QCD results for the deuteron
  binding energy. The first two errors correspond to the ones of the lattice-QCD
results for the binding energy. The last quoted error is our estimation of the uncertainty of the NLO LETs
associated with the unknown $M_\pi$ dependence of the subleading short-range interaction as described in  the text.
\label{predictions}}
\smallskip
\begin{tabular*}{\textwidth}{@{\extracolsep{\fill}}rrrr}
\hline
\hline
\noalign{\smallskip}
  & $M_\pi = 300$ MeV \cite{Yamazaki:2015asa}&     $M_\pi = 390$ MeV \cite{Beane:2011iw} &  $M_\pi = 430$ MeV \cite{Beane:Prelim}
\smallskip
 \\
\hline
\hline \\ [-10pt]
$a M_\pi$ &  $3.59({}^{+0.07}_{-0.05})({}^{+0.07}_{-0.18}) (0.03)$& $4.93({}^{+1.21}_{-0.60})({}^{+\infty}_{-1.09}) ({}^{+0.07}_{-0.09})$   & $4.62({}^{+0.19}_{-0.17})({}^{+0.36}_{-0.27}) ({}^{+0.10}_{-0.14})$  \\  [3.5pt]
$r M_\pi$ & $1.70(0.01)({}^{+0.01}_{-0.04}) (0.03)$ & $2.22({}^{+0.20}_{-0.12})({}^{+1.13}_{-0.25}) ({}^{+0.08}_{-0.10})$  & $2.27({}^{+0.04}_{-0.05})({}^{+0.08}_{-0.07}) ({}^{+0.10}_{-0.14})$  \\  [3.5pt]
$v_2 M_\pi^3$ &  $0.23(0.01)(0.01) ({}^{+0.02}_{-0.01})$  & $0.50({}^{+0.06}_{-0.05})({}^{+0.24}_{-0.10}) ({}^{+0.07}_{-0.09})$ & $0.58({}^{+0.02}_{-0.02})({}^{+0.03}_{-0.03}) ({}^{+0.10}_{-0.13})$  \\  [3.5pt]
$v_3 M_\pi^5$ & $0.11(0.01)(0.01) (0.01)$ &  $0.40({}^{+0.15}_{-0.09})({}^{+0.82}_{-0.16}) ({}^{+0.06}_{-0.07})$  & $0.46({}^{+0.04}_{-0.03})({}^{+0.06}_{-0.05}) (0.12)$ \\  [3.5pt]
$v_4 M_\pi^7$ & $-0.04(0.01) ({}^{+0.03}_{-0.01}) (0.01)$&  $-0.03({}^{+0.05}_{-0.13})({}^{+0.09}_{-0.33}) ({}^{+0.11}_{-0.10})$  &  $0.14({}^{+0.00}_{-0.01})({}^{+0.00}_{-0.01}) ({}^{+0.15}_{-0.16})$    \\
[4pt]
%\smallskip
\hline
\hline
\end{tabular*}
\end{table}
In the case of the preliminary NPLQCD calculation at $M_\pi=430\;$MeV, the
values for the deuteron binding energy and the errors are extracted
from a corresponding plot of  Ref.~\cite{Beane:Prelim}. These
results may serve as useful consistency checks for the already
published lattice-QCD results if the corresponding lattice data
can be used to extract, in
addition to the deuteron binding energy, also information about the scattering parameters.

%%%%%%%%%%%%%%%%%%%%%%%%%%%%%%%%%%%%%%%%%%%%%%%%%%%%%%%%%%%%%%%%%%%%%%%%%%%%%%%%%
\section{Summary and conclusions}
\def\theequation{\arabic{section}.\arabic{equation}}
\label{sec:Summary}

The  quark mass dependence of the low-energy NN scattering  observables
is  investigated  using  the   low-energy  theorems which establish model independent  relations between the coefficients
in the  effective range  expansion governed by the long-range one-pion exchange potential.
In order to clarify the meaning of the LETs,  we employed  the
modified effective range expansion \cite{vanHaeringen:1981pb}  to
parametrize the short-range physics in a systematic way.
Using the knowledge of the long-range interaction together with the MERE
truncated at a given order allows one to
calculate all coefficients of the ERE within a systematically
improvable  expansion in powers of the ratio of the long- and
short-range scales.
The explicit treatment of the left-hand cuts in the partial-wave
scattering amplitudes due to the  long-range pionic forces results in
correlations between the coefficients in the ERE which are regarded as
the LETs.

In this work we considered the LETs for NN scattering in the  $^3S_1$
and $^1S_0$ channels.
At leading order in the above mentioned expansion,  the
short-range physics is parameterized solely in terms of the modified scattering length.
Therefore, in order to obtain LETs at LO,
we considered two nucleons interacting  via the OPEP supplemented  by
two  contact  interactions without derivatives,
whose  strengths were adjusted to reproduce the scattering length  in each of  these  channels.
The NN scattering amplitude was calculated by solving
the Kadyshevsky equation which is exactly renormalizable at this order
\cite{Epelbaum:2012ua}. In the  $^3S_1$  channel,  the predicted values of the effective range
and the shape parameters agree rather well with the empirical values
 extracted from the Nijmegen PWA \cite{deSwart:1995ui}   already  at LO.
On the other hand, the OPEP projected onto the  $^1S_0$ partial wave
is very weak, so that the LETs show much less predictive power in that
channel. We, therefore, restricted ourselves to the spin-triplet
channel in this paper. To further increase the accuracy of our predictions, we
extended the calculations to NLO LETs by including the subleading
short-range interaction and tuning its strength to reproduce the
effective range. The resulting predictions for the shape parameters in
the  $^3S_1$ partial wave
are found to be accurate at the level of a few percent.

As a next step, we generalized  the LETs in the   $^3S_1$  channel to
study correlations between various low-energy observables
at unphysical pion masses up to $M_\pi \simeq 400$ MeV.  To this aim,
we made use of the available lattice-QCD results for the pion mass dependence of
the pion decay constant, nucleon axial vector coupling and the nucleon
mass.  The predicted correlations between the low-energy NN parameters
at unphysical pion masses open the way for nontrivial consistency
checks of the ongoing and upcoming lattice QCD calculations in the NN
sector and might also be useful for reducing  the systematic
uncertainty.

As an application, we extracted  the   low-energy parameters,  namely  the
scattering length,  effective range and  first  three shape parameters,
from the lattice-QCD calculations of the deuteron binding energy at the pion masses
$M_{\pi}= 300$ MeV \cite{Yamazaki:2015asa},  390 MeV \cite{Beane:2011iw}  and
430 MeV \cite{Beane:Prelim}. Our predictions illustrate the importance of
a simultaneous extraction of the scattering parameters in addition to
the binding energies in lattice-QCD simulations of the NN system.

As another application,  we applied the LETs to predict the
$M_{\pi}$ behavior of the binding energy,
scattering length and  shape parameters based on the
linear $M_{\pi}$ dependence of  the effective range $M_\pi r$ suggested
in Ref.~\cite{Beane:2013br}.
We found that the predicted shape of the binding energy as a function
of $M_\pi$ is in  good agreement  with the  general trend  of the
lattice QCD data. We also found that the scattering length and the
shape parameters show rather strong variations with the
pion mass around and below the physical point. This nontrivial
behavior is driven by the long-range physics associated with the
pion exchange and is, in principle, testable in lattice  QCD.

While our analysis establishes correlations between NN low-energy
parameters, it does not provide extrapolation of lattice-QCD results
to the physical point.  The possibility of carrying out chiral extrapolations
of NN scattering observables and the deuteron binding energy by extending the
chiral EFT analysis of Ref.~\cite{Epelbaum:2013ij} beyond the LO
will  be  discussed in a subsequent publication \cite{NNEFT}.

%%%%%%%%%%%%%%%%%%%%%%%%%%%%%%%%%%%%%%%%%%%%%%%%%%%%%%%%%%%%%%%%%%%%%%%%%%%%%%%%%
\section*{Acknowledgments}

We are grateful to Silas Beane, Dean Lee, Ulf-G.~Mei{\ss}ner and
Martin Savage for useful discussions on the topics related to this
work. We are also grateful to Ulf-G.~Mei{\ss}ner for a careful reading
of the manuscript and valuable suggestions.
One of the authors (EE) thanks the organizers of the YITP-T-14-03 workshop on ``Hadrons and Hadron
Interactions in QCD''  at Yukawa Institute for Theoretical
Physics, Kyoto University and the organizers of the Bethe Forum on
Methods in Lattice Field Theory, Bethe Center for Theoretical Physics,
Bonn University,  where a significant part of this work was
carried out.
This work was supported by the European
Community-Research Infrastructure Integrating Activity ``Study of
Strongly Interacting Matter'' (acronym HadronPhysics3,
Grant Agreement n. 283286) under the Seventh Framework Programme of EU,
 the ERC project 259218 NUCLEAREFT, and by the Georgian Shota Rustaveli National Science Foundation (grant 11/31).


\bigskip


\begin{thebibliography}{99}


\bibitem{Bulgac:1997ji}
  A.~Bulgac, G.~A.~Miller and M.~Strikman,
  %``Chiral limit of nuclear physics,''
  Phys.\ Rev.\ C {\bf 56}, 3307 (1997)
  [nucl-th/9708045].
  %%CITATION = NUCL-TH/9708045;%%
  %16 citations counted in INSPIRE as of 18 Mar 2015


\bibitem{Braaten:2003eu}
  E.~Braaten and H.~W.~Hammer,
  %``An Infrared renormalization group limit cycle in QCD,''
  Phys.\ Rev.\ Lett.\  {\bf 91}, 102002 (2003)
  [nucl-th/0303038].
  %%CITATION = NUCL-TH/0303038;%%
  %77 citations counted in INSPIRE as of 18 mar 2015

\bibitem{Epelbaum:2006jc}
  E.~Epelbaum, H.-W.~Hammer, U.-G.~Mei{\ss}ner and A.~Nogga,
  %``More on the infrared renormalization group limit cycle in QCD,''
  Eur.\ Phys.\ J.\ C {\bf 48}, 169 (2006)
  [hep-ph/0602225].
  %%CITATION = HEP-PH/0602225;%%
  %19 citations counted in INSPIRE as of 18 mar 2015


\bibitem{Bedaque:2010hr}
  P.~F.~Bedaque, T.~Luu and L.~Platter,
  %``Quark mass variation constraints from Big Bang nucleosynthesis,''
  Phys.\ Rev.\ C {\bf 83}, 045803 (2011)
  [arXiv:1012.3840 [nucl-th]].
  %%CITATION = ARXIV:1012.3840;%%
  %23 citations counted in INSPIRE as of 18 Mar 2015


\bibitem{Berengut:2013nh}
  J.~C.~Berengut, E.~Epelbaum, V.~V.~Flambaum, C.~Hanhart, U.-G.~Mei{\ss}ner, J.~Nebreda and J.~R.~Pelaez,
  %``Varying the light quark mass: impact on the nuclear force and Big Bang nucleosynthesis,''
  Phys.\ Rev.\ D {\bf 87}, no. 8, 085018 (2013)
  [arXiv:1301.1738 [nucl-th]].
  %%CITATION = ARXIV:1301.1738;%%
  %31 citations counted in INSPIRE as of 18 Mar 2015


\bibitem{Meissner:2014pma}
  U.-G.~Mei{\ss}ner,
  %``Anthropic considerations in nuclear physics,''
  Sci. Bull. (2015) {\bf 60}(1):43-54
  [arXiv:1409.2959 [hep-th]].
  %%CITATION = ARXIV:1409.2959;%%
  %2 citations counted in INSPIRE as of 18 Mar 2015

\bibitem{Hoyle:1954zz}
  F.~Hoyle,
  %``On Nuclear Reactions Occuring in Very Hot Stars. 1. The Synthesis of Elements from Carbon to Nickel,''
  Astrophys.\ J.\ Suppl.\  {\bf 1}, 121 (1954).
  %%CITATION = APJSA,1,121;%%
  %123 citations counted in INSPIRE as of 18 Mar 2015

\bibitem{Oberhummer:2000mn}
  H.~Oberhummer, A.~Csoto and H.~Schlattl,
  %``Bridging the mass gaps at A = 5 and A = 8 in nucleosynthesis,''
  Nucl.\ Phys.\ A {\bf 689}, 269 (2001)
  [nucl-th/0009046].
  %%CITATION = NUCL-TH/0009046;%%
  %19 citations counted in INSPIRE as of 18 Mar 2015

\bibitem{Schlattl:2003dy}
  H.~Schlattl, A.~Heger, H.~Oberhummer, T.~Rauscher and A.~Csoto,
  %``Sensitivity of the c and o production on the 3-alpha rate,''
  Astrophys.\ Space Sci.\  {\bf 291}, 27 (2004)
  [astro-ph/0307528].
  %%CITATION = ASTRO-PH/0307528;%%
  %17 citations counted in INSPIRE as of 18 Mar 2015

\bibitem{Epelbaum:2012iu}
  E.~Epelbaum, H.~Krebs, T.~A.~L\"ahde, D.~Lee and U.-G.~Mei{\ss}ner,
  %``Viability of Carbon-Based Life as a Function of the Light Quark Mass,''
  Phys.\ Rev.\ Lett.\  {\bf 110}, no. 11, 112502 (2013)
  [arXiv:1212.4181 [nucl-th]].
  %%CITATION = ARXIV:1212.4181;%%
  %32 citations counted in INSPIRE as of 18 mar 2015

\bibitem{Epelbaum:2013wla}
  E.~Epelbaum, H.~Krebs, T.~A.~L\"ahde, D.~Lee and U.~G.~Mei{\ss}ner,
  %``Dependence of the triple-alpha process on the fundamental constants of nature,''
  Eur.\ Phys.\ J.\ A {\bf 49}, 82 (2013)
  [arXiv:1303.4856 [nucl-th]].
  %%CITATION = ARXIV:1303.4856;%%
  %18 citations counted in INSPIRE as of 18 Mar 2015

\bibitem{Lee:2008fa}
  D.~Lee,
  %``Lattice simulations for few- and many-body systems,''
  Prog.\ Part.\ Nucl.\ Phys.\  {\bf 63}, 117 (2009)
  [arXiv:0804.3501 [nucl-th]].
  %%CITATION = ARXIV:0804.3501;%%
  %86 citations counted in INSPIRE as of 18 mar 2015

\bibitem{Epelbaum:2008ga}
  E.~Epelbaum, H.~W.~Hammer and U.-G.~Mei{\ss}ner,
  %``Modern Theory of Nuclear Forces,''
  Rev.\ Mod.\ Phys.\  {\bf 81}, 1773 (2009)
  [arXiv:0811.1338 [nucl-th]].
  %%CITATION = ARXIV:0811.1338;%%
  %463 citations counted in INSPIRE as of 18 Mar 2015

\bibitem{Machleidt:2011zz}
  R.~Machleidt and D.~R.~Entem,
  %``Chiral effective field theory and nuclear forces,''
  Phys.\ Rept.\  {\bf 503}, 1 (2011)
  [arXiv:1105.2919 [nucl-th]].
  %%CITATION = ARXIV:1105.2919;%%
  %266 citations counted in INSPIRE as of 18 Mar 2015

\bibitem{Epelbaum:2014sza}
  E.~Epelbaum, H.~Krebs and U.-G.~Mei{\ss}ner,
  %``Precision nucleon-nucleon potential at fifth order in the chiral expansion,''
  arXiv:1412.4623 [nucl-th].
  %%CITATION = ARXIV:1412.4623;%%
  %4 citations counted in INSPIRE as of 18 mar 2015

\bibitem{Entem:2014msa}
  D.~R.~Entem, N.~Kaiser, R.~Machleidt and Y.~Nosyk,
  %``Peripheral nucleon-nucleon scattering at fifth order of chiral perturbation theory,''
  Phys.\ Rev.\ C {\bf 91}, no. 1, 014002 (2015)
  [arXiv:1411.5335 [nucl-th]].
  %%CITATION = ARXIV:1411.5335;%%
  %3 citations counted in INSPIRE as of 31 Mar 2015

\bibitem{Epelbaum:2014efa}
  E.~Epelbaum, H.~Krebs and U.-G.~Mei{\ss}ner,
  %``Improved chiral nucleon-nucleon potential up to next-to-next-to-next-to-leading order,''
  arXiv:1412.0142 [nucl-th].
  %%CITATION = ARXIV:1412.0142;%%
  %10 citations counted in INSPIRE as of 18 Mar 2015

\bibitem{Beane:2001bc}
  S.~R.~Beane, P.~F.~Bedaque, M.~J.~Savage and U.~van Kolck,
  %``Towards a perturbative theory of nuclear forces,''
  Nucl.\ Phys.\ A {\bf 700}, 377 (2002)
  [nucl-th/0104030].
  %%CITATION = NUCL-TH/0104030;%%
  %223 citations counted in INSPIRE as of 18 Mar 2015

\bibitem{Beane:2002xf}
  S.~R.~Beane and M.~J.~Savage,
  %``The Quark mass dependence of two nucleon systems,''
  Nucl.\ Phys.\ A {\bf 717}, 91 (2003)
  [nucl-th/0208021].
  %%CITATION = NUCL-TH/0208021;%%
  %112 citations counted in INSPIRE as of 18 mar 2015

\bibitem{Epelbaum:2002gb}
  E.~Epelbaum, U.-G.~Mei{\ss}ner and W.~Gl\"ockle,
  %``Nuclear forces in the chiral limit,''
  Nucl.\ Phys.\ A {\bf 714}, 535 (2003)
  [nucl-th/0207089].
  %%CITATION = NUCL-TH/0207089;%%
  %142 citations counted in INSPIRE as of 18 Mar 2015

\bibitem{Beane:2002vs}
  S.~R.~Beane and M.~J.~Savage,
  %``Variation of fundamental couplings and nuclear forces,''
  Nucl.\ Phys.\ A {\bf 713}, 148 (2003)
  [hep-ph/0206113].
  %%CITATION = HEP-PH/0206113;%%
  %103 citations counted in INSPIRE as of 18 Mar 2015

\bibitem{Epelbaum:2002gk}
  E.~Epelbaum, U.-G.~Mei{\ss}ner and W.~Gl\"ockle,
  %``Further comments on nuclear forces in the chiral limit,''
  nucl-th/0208040.
  %%CITATION = NUCL-TH/0208040;%%
  %27 citations counted in INSPIRE as of 18 Mar 2015

\bibitem{Soto:2011tb}
  J.~Soto and J.~Tarrus,
  %``On the quark mass dependence of nucleon-nucleon S-wave scattering lengths,''
  Phys.\ Rev.\ C {\bf 85}, 044001 (2012)
  [arXiv:1112.4426 [nucl-th]].
  %%CITATION = ARXIV:1112.4426;%%
  %13 citations counted in INSPIRE as of 18 Mar 2015

\bibitem{Barnea:2013uqa}
  N.~Barnea, L.~Contessi, D.~Gazit, F.~Pederiva and U.~van Kolck,
  %``Effective Field Theory for Lattice Nuclei,''
  Phys.\ Rev.\ Lett.\  {\bf 114}, no. 5, 052501 (2015)
  [arXiv:1311.4966 [nucl-th]].
  %%CITATION = ARXIV:1311.4966;%%
  %4 citations counted in INSPIRE as of 18 mar 2015

\bibitem{Flambaum:2007mj}
  V.~V.~Flambaum and R.~B.~Wiringa,
  %``Dependence of nuclear binding on hadronic mass variation,''
  Phys.\ Rev.\ C {\bf 76}, 054002 (2007)
  [arXiv:0709.0077 [nucl-th]].
  %%CITATION = ARXIV:0709.0077;%%
  %55 citations counted in INSPIRE as of 18 Mar 2015

\bibitem{Mondejar:2006yu}
  J.~Mondejar and J.~Soto,
  %``The nucleon-nucleon potential beyond the static approximation,''
  Eur.\ Phys.\ J.\ A {\bf 32}, 77 (2007)
  [nucl-th/0612051].
  %%CITATION = NUCL-TH/0612051;%%
  %18 citations counted in INSPIRE as of 18 mar 2015

\bibitem{Kaplan:1998we}
  D.~B.~Kaplan, M.~J.~Savage and M.~B.~Wise,
  %``Two nucleon systems from effective field theory,''
  Nucl.\ Phys.\ B {\bf 534}, 329 (1998)
  [nucl-th/9802075].
  %%CITATION = NUCL-TH/9802075;%%
  %467 citations counted in INSPIRE as of 19 mar 2015

\bibitem{Cohen:1998jr}
  T.~D.~Cohen and J.~M.~Hansen,
  %``Low-energy theorems for nucleon-nucleon scattering,''
  Phys.\ Rev.\ C {\bf 59}, 13 (1999) [nucl-th/9808038];
   %``Testing low-energy theorems in nucleon-nucleon scattering,''
  Phys.\ Rev.\ C {\bf 59}, 3047 (1999)
  [nucl-th/9901065].
%  ;
  %%CITATION = NUCL-TH/9808038;%%
  %76 citations counted in INSPIRE as of 19 Mar 2015

\bibitem{Fleming:1999ee}
  S.~Fleming, T.~Mehen and I.~W.~Stewart,
  %``NNLO corrections to nucleon-nucleon scattering and perturbative pions,''
  Nucl.\ Phys.\ A {\bf 677}, 313 (2000)
  [nucl-th/9911001].
  %%CITATION = NUCL-TH/9911001;%%
  %140 citations counted in INSPIRE as of 19 mar 2015

\bibitem{ressat}
  E.~Epelbaum, U.-G.-Mei{\ss}ner, W.~Gl\"ockle, C.~Elster,
  %``Resonance saturation for four nucleon operators,''
  Phys.\ Rev.\ C {\bf 65}, 044001 (2002).

\bibitem{Epelbaum:2013ij}
  E.~Epelbaum and J.~Gegelia,
  %``The two-nucleon problem in EFT reformulated: Pion and nucleon masses as soft and hard scales,''
  PoS CD {\bf 12}, 090 (2013)
  [arXiv:1301.6134 [nucl-th]].
  %%CITATION = ARXIV:1301.6134;%%
  %5 citations counted in INSPIRE as of 19 Mar 2015

\bibitem{Epelbaum:2012ua}
  E.~Epelbaum and J.~Gegelia,
  %``Weinberg's approach to nucleon-nucleon scattering revisited,''
  Phys.\ Lett.\ B {\bf 716}, 338 (2012)
  [arXiv:1207.2420 [nucl-th]].
  %%CITATION = ARXIV:1207.2420;%%
  %23 citations counted in INSPIRE as of 19 Mar 2015

\bibitem{Kadyshevsky:1967rs}
  V.~G.~Kadyshevsky,
  %``Quasipotential type equation for the relativistic scattering amplitude,''
  Nucl.\ Phys.\ B {\bf 6}, 125 (1968).
  %%CITATION = NUPHA,B6,125;%%
  %224 citations counted in INSPIRE as of 20 mar 2015

\bibitem{Fukugita:1994ve}
  M.~Fukugita, Y.~Kuramashi, M.~Okawa, H.~Mino and A.~Ukawa,
  %``Hadron scattering lengths in lattice QCD,''
  Phys.\ Rev.\ D {\bf 52}, 3003 (1995)
  [hep-lat/9501024].
  %%CITATION = HEP-LAT/9501024;%%
  %200 citations counted in INSPIRE as of 20 mar 2015

\bibitem{Beane:2006mx}
  S.~R.~Beane, P.~F.~Bedaque, K.~Orginos and M.~J.~Savage,
  %``Nucleon-nucleon scattering from fully-dynamical lattice QCD,''
  Phys.\ Rev.\ Lett.\  {\bf 97}, 012001 (2006)
  [hep-lat/0602010].
  %%CITATION = HEP-LAT/0602010;%%
  %166 citations counted in INSPIRE as of 20 mar 2015

\bibitem{Beane:2011iw}
  S.~R.~Beane {\it et al.}  [NPLQCD Collaboration],
  %``The Deuteron and Exotic Two-Body Bound States from Lattice QCD,''
  Phys.\ Rev.\ D {\bf 85}, 054511 (2012)
  [arXiv:1109.2889 [hep-lat]].
  %%CITATION = ARXIV:1109.2889;%%
  %73 citations counted in INSPIRE as of 04 mar 2015

\bibitem{Beane:2012vq}
  S.~R.~Beane {\it et al.}  [NPLQCD Collaboration],
  %``Light Nuclei and Hypernuclei from Quantum Chromodynamics in the Limit of SU(3) Flavor Symmetry,''
  Phys.\ Rev.\ D {\bf 87}, no. 3, 034506 (2013)
  [arXiv:1206.5219 [hep-lat]].
  %%CITATION = ARXIV:1206.5219;%%
  %68 citations counted in INSPIRE as of 20 mar 2015

\bibitem{Yamazaki:2012hi}
  T.~Yamazaki, K.~i.~Ishikawa, Y.~Kuramashi and A.~Ukawa,
  %``Helium nuclei, deuteron and dineutron in 2+1 flavor lattice QCD,''
  Phys.\ Rev.\ D {\bf 86}, 074514 (2012)
  [arXiv:1207.4277 [hep-lat]].
  %%CITATION = ARXIV:1207.4277;%%
  %45 citations counted in INSPIRE as of 04 mar 2015

\bibitem{Yamazaki:2009ua}
  T.~Yamazaki {\it et al.}  [PACS-CS Collaboration],
  %``Helium Nuclei in Quenched Lattice QCD,''
  Phys.\ Rev.\ D {\bf 81}, 111504 (2010)
  [arXiv:0912.1383 [hep-lat]].
  %%CITATION = ARXIV:0912.1383;%%
  %66 citations counted in INSPIRE as of 20 mar 2015

\bibitem{Beane:2013br}
  S.~R.~Beane {\it et al.}  [NPLQCD Collaboration],
  %``Nucleon-Nucleon Scattering Parameters in the Limit of SU(3) Flavor Symmetry,''
  Phys.\ Rev.\ C {\bf 88}, no. 2, 024003 (2013)
  [arXiv:1301.5790 [hep-lat]].
  %%CITATION = ARXIV:1301.5790;%%
  %29 citations counted in INSPIRE as of 04 mar 2015

  \bibitem{Inoue:2011ai}
  T.~Inoue {\it et al.}  [HAL QCD Collaboration],
  %``Two-Baryon Potentials and H-Dibaryon from 3-flavor Lattice QCD Simulations,''
  Nucl.\ Phys.\ A {\bf 881}, 28 (2012)
  [arXiv:1112.5926 [hep-lat]].
  %%CITATION = ARXIV:1112.5926;%%

\bibitem{Yamazaki:2015asa}
  T.~Yamazaki, K.~i.~Ishikawa, Y.~Kuramashi and A.~Ukawa,
  %``Study of quark mass dependence of binding energy for light nuclei in 2+1 flavor lattice QCD,''
  arXiv:1502.04182 [hep-lat].
  %%CITATION = ARXIV:1502.04182;%%

\bibitem{Beane:Prelim}
  S.~R.~Beane, \emph{Nuclear Structure from First Principles}, talk at the
HHIQCD Workshop, Feb 15 - Mar 21, 2015, Yukawa Institute for Theoretical Physics, Kyoto University, Japan

\bibitem{Steele:1998zc}
  J.~V.~Steele and R.~J.~Furnstahl,
  %``Removing pions from two nucleon effective field theory,''
  Nucl.\ Phys.\ A {\bf 645}, 439 (1999)
  [nucl-th/9808022].
  %%CITATION = NUCL-TH/9808022;%%
  %46 citations counted in INSPIRE as of 30 Mar 2015

\bibitem{Epelbaum:2009sd}
  E.~Epelbaum and J.~Gegelia,
  %``Regularization, renormalization and 'peratization' in effective field theory for two nucleons,''
  Eur.\ Phys.\ J.\ A {\bf 41}, 341 (2009)
  [arXiv:0906.3822 [nucl-th]].
  %%CITATION = ARXIV:0906.3822;%%
  %66 citations counted in INSPIRE as of 30 mar 2015

\bibitem{Epelbaum:2009zz}
  E.~Epelbaum and J.~Gegelia,
  %``Effective field theory for nuclear forces,''
  PoS CD {\bf 09}, 077 (2009).
  %%CITATION = POSCI,CD09,077;%%
  %3 citations counted in INSPIRE as of 30 Mar 2015

\bibitem{Epelbaum:2010nr}
  E.~Epelbaum,
  %``Nuclear Forces from Chiral Effective Field Theory: A Primer,''
  arXiv:1001.3229 [nucl-th].
  %%CITATION = ARXIV:1001.3229;%%
  %10 citations counted in INSPIRE as of 30 Mar 2015

\bibitem{Gasparyan:2012km}
  A.~M.~Gasparyan, M.~F.~M.~Lutz and E.~Epelbaum,
  %``Two-nucleon scattering: Merging chiral effective field theory with dispersion relations,''
  Eur.\ Phys.\ J.\ A {\bf 49}, 115 (2013)
  [arXiv:1212.3057 [nucl-th]].
  %%CITATION = ARXIV:1212.3057;%%
  %14 citations counted in INSPIRE as of 30 Mar 2015

\bibitem{Oller:2014uxa}
  J.~A.~Oller,
  %``Nucleon-Nucleon scattering from dispersion relations: next-to-next-to-leading order study,''
  arXiv:1402.2449 [nucl-th].
  %%CITATION = ARXIV:1402.2449;%%
  %1 citations counted in INSPIRE as of 30 Mar 2015

\bibitem{Blatt:49}
J.~M. Blatt and J.~D. Jackson,
\newblock Phys. Rev. {\bf 76}, 18 (1949).

\bibitem{Bethe:49}
H.~A. Bethe,
\newblock Phys. Rev. {\bf 76}, 38 (1949).

\bibitem{Midya:2015eta}
  B.~Midya, J.~Evrard, S.~Abramowicz, O.~L.~R.~Su\'arez and J.~M.~Sparenberg,
  %``Supersymmetric inversion of effective-range expansions,''
  arXiv:1501.04011 [quant-ph].
  %%CITATION = ARXIV:1501.04011;%%

\bibitem{PavonValderrama:2005ku}
  M.~Pavon Valderrama and E.~Ruiz Arriola,
  %``Low-energy NN scattering at next-to-next-to-next-to-next-to-leading order for partial waves with j [<LESS-THAN OR EQUAL TO>] 5,''
  Phys.\ Rev.\ C {\bf 72}, 044007 (2005).
  %%CITATION = PHRVA,C72,044007;%%

\bibitem{vanHaeringen:1981pb}
H.~van Haeringen and L.~P. Kok,
\newblock Phys. Rev. {\bf A26}, 1218 (1982).
%%CITATION = PHRVA,A26,1218;%%

\bibitem{Badalian:1981xj}
A.~M. Badalian, L.~P. Kok, M.~I. Polikarpov, and Y.~A. Simonov,
\newblock Phys. Rept. {\bf 82}, 31 (1982).
%%CITATION = PRPLC,82,31;%%

\bibitem{Epelbaum:2006pt}
  E.~Epelbaum and U.-G.~Mei{\ss}ner,
  %``On the Renormalization of the One-Pion Exchange Potential and the Consistency of Weinberg`s Power Counting,''
  Few Body Syst.\  {\bf 54}, 2175 (2013)
  [nucl-th/0609037].
  %%CITATION = NUCL-TH/0609037;%%
  %93 citations counted in INSPIRE as of 31 mar 2015

\bibitem{Gezerlis:2013ipa}
  A.~Gezerlis, I.~Tews, E.~Epelbaum, S.~Gandolfi, K.~Hebeler, A.~Nogga and A.~Schwenk,
  %``Quantum Monte Carlo Calculations with Chiral Effective Field Theory Interactions,''
  Phys.\ Rev.\ Lett.\  {\bf 111}, no. 3, 032501 (2013)
  [arXiv:1303.6243 [nucl-th]].
  %%CITATION = ARXIV:1303.6243;%%
  %54 citations counted in INSPIRE as of 31 Mar 2015

\bibitem{Gezerlis:2014zia}
  A.~Gezerlis, I.~Tews, E.~Epelbaum, M.~Freunek, S.~Gandolfi, K.~Hebeler, A.~Nogga and A.~Schwenk,
  %``Local chiral effective field theory interactions and quantum Monte Carlo applications,''
  Phys.\ Rev.\ C {\bf 90}, no. 5, 054323 (2014)
  [arXiv:1406.0454 [nucl-th]].
  %%CITATION = ARXIV:1406.0454;%%
  %14 citations counted in INSPIRE as of 31 Mar 2015

\bibitem{Lepage:1997cs}
  G.~P.~Lepage,
  %``How to renormalize the Schrodinger equation,''
  nucl-th/9706029.
  %%CITATION = NUCL-TH/9706029;%%
  %310 citations counted in INSPIRE as of 31 Mar 2015

\bibitem{NNEFT}
  V.~Baru, E.~Epelbaum, A.~A.~Filin,  and  J.~Gegelia,
  ``Chiral extrapolations of the NN  low-energy parameters,''
 in preparation.

\bibitem{Stoks:1994wp}
  V.~G.~J.~Stoks, R.~A.~M.~Klomp, C.~P.~F.~Terheggen and J.~J.~de Swart,
  %``Construction of high quality N N potential models,''
  Phys.\ Rev.\ C {\bf 49}, 2950 (1994)
  [nucl-th/9406039].
  %%CITATION = NUCL-TH/9406039;%%

\bibitem{Blatt:1952zz}
  J.~M.~Blatt and L.~C.~Biedenharn,
  %``The Angular Distribution of Scattering and Reaction Cross Sections,''
  Rev.\ Mod.\ Phys.\  {\bf 24}, 258 (1952).
  %%CITATION = RMPHA,24,258;%%
  %121 citations counted in INSPIRE as of 31 Mar 2015

\bibitem{Epelbaum:2015sha}
  E.~Epelbaum, A.~M.~Gasparyan, J.~Gegelia and H.~Krebs,
  %``$^1S_0$ nucleon-nucleon scattering in the modified Weinberg approach,''
  arXiv:1501.01191 [nucl-th].
  %%CITATION = ARXIV:1501.01191;%%{PavonValderrama:2005wv}

\bibitem{deSwart:1995ui}
  J.~J.~de Swart, C.~P.~F.~Terheggen and V.~G.~J.~Stoks,
  %``The Low-energy n p scattering parameters and the deuteron,''
  nucl-th/9509032.
  %%CITATION = NUCL-TH/9509032;%%
  %88 citations counted in INSPIRE as of 31 Mar 2015

\bibitem{PavonValderrama:2005wv}
  M.~Pavon Valderrama and E.~Ruiz Arriola,
  %``Renormalization of NN interaction with chiral two pion exchange potential. central phases and the deuteron,''
  Phys.\ Rev.\ C {\bf 74}, 054001 (2006)
  [nucl-th/0506047].
  %%CITATION = NUCL-TH/0506047;%%

\bibitem{Elhatisari:2013swa}
  S.~Elhatisari, S.~K\"onig, D.~Lee and H.-W.~Hammer,
  %``Causality, universality, and effective field theory for van der Waals interactions,''
  Phys.\ Rev.\ A {\bf 87}, no. 5, 052705 (2013)
  [arXiv:1303.5261 [physics.atom-ph]].

\bibitem{Durr:2013goa}
  S.~D\"urr {\it et al.}  [Budapest-Marseille-Wuppertal Collaboration],
  %``Lattice QCD at the physical point meets SU(2) chiral perturbation theory,''
  Phys.\ Rev.\ D {\bf 90}, no. 11, 114504 (2014)
  [arXiv:1310.3626 [hep-lat]].
  %%CITATION = ARXIV:1310.3626;%%
  %13 citations counted in INSPIRE as of 24 Feb 2015

\bibitem{Alexandrou:2013jsa}
  C.~Alexandrou, M.~Constantinou, V.~Drach, K.~Jansen, C.~Kallidonis and G.~Koutsou,
  %``Nucleon generalized form factors with twisted mass fermions,''
  PoS LATTICE {\bf 2013}, 292 (2014)
  [arXiv:1312.2874 [hep-lat]].
  %%CITATION = ARXIV:1312.2874;%%
  %11 citations counted in INSPIRE as of 24 Feb 2015

\bibitem{Alexandrou:2014wca}
  C.~Alexandrou, M.~Constantinou, K.~Hadjiyiannakou, K.~Jansen, C.~Kallidonis and G.~Koutsou,
  %``Nucleon observables and axial charges of other baryons using twisted mass fermions,''
  arXiv:1411.3494 [hep-lat].
  %%CITATION = ARXIV:1411.3494;%%
  %1 citations counted in INSPIRE as of 24 Feb 2015

\bibitem{Horsley:2013ayv}
  R.~Horsley, Y.~Nakamura, A.~Nobile, P.~E.~L.~Rakow, G.~Schierholz and J.~M.~Zanotti,
  %``Nucleon axial charge and pion decay constant from two-flavor lattice QCD,''
  Phys.\ Lett.\ B {\bf 732}, 41 (2014)
  [arXiv:1302.2233 [hep-lat]].
  %%CITATION = ARXIV:1302.2233;%%
  %22 citations counted in INSPIRE as of 24 Apr 2015

\end{thebibliography}
\end{document}